\documentclass[11pt]{article}

\usepackage{amsmath}
\usepackage{authblk}
\usepackage{enumitem}  
\usepackage{hyperref}  
\usepackage[usenames, dvipsnames]{color} 
\usepackage{graphicx}     
\usepackage{subcaption}
\usepackage[utf8]{inputenc}
\usepackage{url}
\usepackage{cite}
\usepackage{amssymb}
\usepackage{mathrsfs}

\usepackage[dvipsnames]{xcolor}
\usepackage{physics}

\usepackage{overpic}

\usepackage{hyperref}

\hypersetup{
    bookmarks=true,         
    unicode=false,          
    pdftoolbar=true,        
    pdfmenubar=true,        
    pdffitwindow=false,     
    pdfstartview={FitH},    
    pdfauthor={Author},     
    pdfsubject={Subject},   
    pdfcreator={Creator},   
    pdfproducer={Producer}, 
    pdfkeywords={keyword1} {key2} {key3}, 
    pdfnewwindow=true,      
    linktocpage=true,
    linkcolor=blue,           
     linkbordercolor=blue,
    citecolor=blue,        
    filecolor=blue,      
    urlcolor=blue         
} 

\def\beq{\begin{equation}}
\def\eeq{\end{equation}}

\newcommand{\ben}{\begin{enumerate}}
\newcommand{\een}{\end{enumerate}}
\newcommand{\be}{\begin{equation}}
\newcommand{\ee}{\end{equation}}

\usepackage{tikz}
\usetikzlibrary{positioning}
\usetikzlibrary{intersections}
\usetikzlibrary{fadings} 
\usetikzlibrary{arrows.meta} 
\usetikzlibrary{arrows}

\tikzfading[name=fade out,
inner color=transparent!0,
outer color=transparent!100]

\definecolor{cherryblossompink}{rgb}{1.0, 0.72, 0.77}
\definecolor{lightblue}{rgb}{0.68, 0.85, 0.9}

\usetikzlibrary{decorations.pathmorphing}
\usetikzlibrary{decorations.pathreplacing,decorations.markings}

\usetikzlibrary{backgrounds,automata}

\begin{document}
 
\numberwithin{equation}{section}
 
 \title{\vspace{-2cm}\bf\LARGE  The minus sign in the first law\\ of 
de Sitter horizons \\[9mm]
}

\author[1]{\large Batoul Banihashemi\thanks{baniha@umd.edu}}
\author[1,4]{\large Ted Jacobson\thanks{jacobson@umd.edu}}
\author[2]{\large Andrew Svesko\thanks{a.svesko@ucl.ac.uk}}
\author[3,4]{\large Manus    Visser\thanks{mv551@cam.ac.uk}} 

\vspace{10mm}

\affil[1]{\it{\small Maryland Center for Fundamental Physics, University of Maryland,

College Park, MD 20742, USA}}

\affil[2]{\it{\small Department of Physics and Astronomy, University College London,

 London, WC1E 6BT, United Kingdom}} 
\affil[3]{\it{\small Department of Theoretical Physics,  University of Geneva,  

24 quai Ernest-Ansermet, 1211 Gen\`eve 4, Switzerland}}
\affil[4]{\it{\small
Department of Applied Mathematics and Theoretical Physics, University of Cambridge,
Wilberforce Road, Cambridge CB3 0WA, UK}}
 
 \date{ }
 
\maketitle

\begin{abstract}
\noindent Due to a well-known, but curious, minus sign in the Gibbons-Hawking first law for the static patch of de Sitter space, the entropy of the cosmological horizon   is {\it reduced} by the addition of Killing energy. This minus sign raises the puzzling question how the thermodynamics of the static patch should be understood. We argue the confusion arises because of a mistaken interpretation of the matter Killing energy as the total internal energy, and resolve the puzzle by introducing a  system boundary at which a proper thermodynamic ensemble can be specified. When this boundary shrinks to zero size the total internal energy of the ensemble (the Brown-York energy) vanishes, as does its variation.
Part of this vanishing variation is thermalized, captured by the horizon entropy variation, and part is the matter contribution, which may or may not be thermalized. If the matter is in global equilibrium at the de Sitter temperature, the first law becomes the statement that the generalized  entropy is stationary. 
  
  \end{abstract}

\thispagestyle{empty}

\newpage

 \tableofcontents
 
\section{Introduction} \label{sec:intro}

Shortly after the discovery of the laws of black hole mechanics \cite{Bekenstein:1973ur,Bardeen:1973gs} and the Hawking effect \cite{Hawking:1975vcx}, Gibbons and Hawking (GH) \cite{Gibbons:1977mu}
found that those black hole phenomena extend quite naturally to the 
case of ``cosmological horizons". One of the many results they established is 
a ``first law of event horizons", which relates variations away from 
a given Kerr-de Sitter (KdS) spacetime:
\begin{equation}
    \int_\Sigma \delta T_{\mu\nu}\xi^\mu d\Sigma^\nu = -\kappa_c \delta A_c/8\pi G 
    -\kappa_b \delta A_b/8\pi G - \Omega_b\delta J_b \, . \label{GHfirstlaw}
\end{equation}
Here the integral is over a spatial slice $\Sigma$ bounded by the cosmological and black hole horizons, $T_{\mu\nu}$~is the matter energy momentum tensor, $\xi^\mu$ is the 
Killing vector that generates the cosmological KdS horizon, the subscripts
$c$ and $b$ on the  (positive) surface gravities $\kappa$ and areas $A$ refer to the cosmological and black hole horizons, and $\Omega_b$ and $J_b$ are the angular velocity and angular momentum of the 
black hole relative to the cosmological horizon.\footnote{In this paper we use metric signature $(-+++)$ and   set the speed of light equal to one.}$^{,}$\footnote{See also \cite{Dolan:2013ft} for a derivation of \eqref{GHfirstlaw}, without the stress-energy term,  using the Hamiltonian formalism.}  
In the case when there is no black hole present, the GH first law of event horizons reduces to a statement about variations away from a static patch of de Sitter (dS) spacetime, 
\begin{equation}\label{GH2}
    \int_\Sigma \delta T_{\mu\nu}\xi^\mu d\Sigma^\nu = -\kappa\, \delta A/8\pi G \, , 
\end{equation}
where the subscript $c$ is now dropped since only one horizon is present.

Another result of Gibbons and Hawking in the same paper
is that, when restricted 
to a static patch, the de Sitter vacuum state of quantum fields 
is thermal with respect to
the Hamiltonian generating the Killing flow, at temperature $T_{\rm GH}=\hbar \kappa/2\pi$.\footnote{This definition of the GH temperature is proportional to the normalization of $\xi^\mu$.
If $\xi^\mu$ is normalized to unity at the center of the static patch
it agrees with the standard definition, $\hbar/2\pi L$, where $L$ is
the de Sitter length.} Together with the Bekenstein-Hawking entropy formula $S_{\rm BH}= A/4 G\hbar$,    
this implies that 
the right-hand side of \eqref{GH2} takes on a thermodynamic
form, $-T_{\rm GH} \delta S_{\rm BH}$. It follows that the entropy of the static patch is {\it reduced} by the addition of Killing
energy. This decrease of entropy is opposite to the 
usual expectation when energy is added to a 
positive temperature thermal state, and this opposite
sign has raised the puzzling
question of how the thermodynamics of a static patch should be understood. 
Despite the long time that has passed since the original
observation by Gibbons and Hawking, 
in our view a satisfactory answer to this
puzzling question has not yet been given.

Several proposed thermodynamic interpretations of 
the minus sign 
have appeared in the literature.
In their pioneering paper, GH
considered a physical process in which
a ``particle" detector in the static patch
absorbs a quantum from the perceived 
thermal bath of vacuum fluctuations,
and then the system settles down to a
new stationary state in which, according to
\eqref{GH2}, the increase of detector energy
is accompanied by a decrease of horizon area.
Their take was that
``One can interpret this as a reduction
in the entropy of the universe beyond the
event horizon caused by the propagation of some
radiation from this region to the observer''.
In this view, the entropy is attributed not to
the static patch itself, but to the region 
outside the horizon. It is not clear to us 
what to make of this reasoning. For one thing,
the notion that the entropy refers to degrees of freedom 
behind the horizon, although popular especially in the context of black hole physics, seems inconsistent with the thermodynamic notion of the entropy of a system that interacts directly with agents \cite{Sorkin:1997ja}. 
Moreover, if the total state is pure, then the entropy outside
the horizon would presumably be equal to the entropy inside the horizon.
When referred to that entropy, the same argument would imply that it should
have increased rather than decreased. 

Over the intervening years, variations on the GH interpretation have been
discussed in the literature. 
For instance, Ref.~\cite{Spradlin:2001pw} 
also took the viewpoint 
that the entropy refers to the
region outside the horizon, and 
posited that therefore the energy relevant 
to the first law should be the energy behind the horizon. 
It was further argued that, since 
the total matter Killing energy variation 
on a closed slice of global de Sitter space 
vanishes,\footnote{In Ref.\cite{Spradlin:2001pw} this 
vanishing was explained verbally and
illustrated by an example in three spacetime dimensions.
Since we are not aware of an explicit demonstration in the
literature, we have included one in Appendix \ref{app:zeroKillingenergy}.}
the
Killing energy variation behind the horizon is the negative of the left-hand
side of~\eqref{GH2}. In this way, \eqref{GH2} is 
read as an instance of the ordinary first law applied to the region behind the horizon. It seems to us this proposed interpretation has 
a serious drawback in addition to 
the point above regarding which degrees of freedom the
entropy refers to. Namely, although
the Killing energy outside the horizon is the negative of that inside, 
the Killing vector is {\it past pointing} outside if it is future pointing inside, which means this ``energy'' is the opposite of what would normally be thermodynamically relevant. Another variation of this idea is expressed in \cite{Dinsmore:2019elr}, which
attributes the minus sign to the fact that energy as ``drawn from a de Sitter bath of energy, reducing the entropy of the bath''.  But it is not clear why, when comparing two solutions, energy 
of a black hole (or other object)   in the static patch
must be regarded as having been taken away from the de Sitter bath.

An information-theoretic, physical process 
interpretation of a reversed relation was offered in 
\cite{Anninos:2012qw}, in direct analogy with the
case of a black hole: 
``The entropy increases when we throw mass outside the cosmological horizon
surrounding us. Indeed, the less information we have about the interior of the cosmological horizon, the higher its entropy will be."
Something along these lines makes sense to us but, as explained above
we think it is more correct to attribute the entropy
to the region accessible to the observers who assign this
entropy, i.e., to the region
{\it inside} of the static patch.
When it crosses the horizon,\footnote{Strictly speaking, 
a patch observer never sees it cross the horizon, 
but does see it cross the “stretched horizon”.} 
the matter energy is ``thermalized" from
the viewpoint of these observers, and
the entropy of the state accessible 
to them thus increases.
We shall discuss this interpretation further below in section~\ref{sec:physicalprocess}, 
where we also consider the
case of adding mass by throwing it across the past horizon
into the observers' patch. 
 

A rather different proposed interpretation of the minus sign 
is that it occurs because the temperature of the static patch has in fact the {\it negative} value $-T_{\rm GH}$
 \cite{Klemm:2004mb}. 
A system can have negative temperature only if its Hamiltonian is 
bounded above, and it was pointed out in \cite{Klemm:2004mb} that 
this could make sense for the static patch, since there is a largest mass
for a  de Sitter black hole,
and because there is reason to 
believe that the static patch must be described by a finite dimensional
 Hilbert space \cite{Banks:2000fe,Fischler}. A major difficulty with this proposal, however, is that the Killing temperature of quantum fields in the static patch is known to be {\it positive} (and equal to $T_{\rm GH}$).
This proposal was endorsed in \cite{Jacobson:2018ahi,Jacobson:2019gco} (in the context of more general causal diamonds in maximally symmetric spacetimes), but we no longer think that arguments 
given there for the negative temperature interpretation are cogent.

Finally, a statistical explanation of the minus sign was given in \cite{Banks:2006rx} (see also \cite{Banks:2013fr,Banks:2018jqo} and \cite{Dinsmore:2019elr}). It was 
  argued there in the context of a matrix model
that, since the GH entropy of ``de Sitter space" corresponds to
the entropy of the maximally mixed state for a region surrounded by a cosmological horizon,  where nothing is fixed but the topology \cite{Banks:2000fe},\!\!\cite{Banks:2003ta},\!\!\cite{Banks:2006rx}  
(see also  \cite{Dong:2018cuv,Banihashemi:2022jys,Chandrasekaran:2022cip}),
any further specification of the state, such as the presence of a black hole or a certain amount of matter Killing energy, amounts to a ``constraint" on the state. This results in a smaller entropy than that of the maximally mixed state. 
One of the three arguments in \cite{Banks:2000fe} for this maximally mixed state interpretation of the Gibbons-Hawking entropy of ``de Sitter space" was based on the complete diffeomorphism invariance of the GH partition function in the absence of boundary conditions. In \cite{Banihashemi:2022jys} this reasoning is supported by viewing this state as the limit of a canonical ensemble defined by conditions at a York boundary that shrinks to zero inside the horizon.
This statistical interpretation of the minus sign is fundamental, 
but it does not directly clarify the thermodynamic interpretation
of the first law.

In this paper we shall argue that 
the puzzle of the minus sign is resolved by 
sharpening the definition of 
the thermodynamic ensemble to which
the first law is applied, and by clarifying the 
thermodynamic meaning of the so-called first law. 
To specify the thermodynamic ensemble we introduce a system boundary, 
inside the static patch, at which the temperature of the ensemble can be 
specified. This is a de Sitter version of the approach York and collaborators
have taken to defining  quasi-local thermodynamic ensembles containing black holes \cite{York:1986it, Whiting:1988qr, Braden:1990hw,Brown:1992br,Brown:1992bq,Brown:1994gs},
and it has also been applied to the case of cosmological
horizons \cite{Hayward:1990zm,Miyashita:2021iru,Svesko:2022txo,Draper:2022ofa,Banihashemi:2022jys}. 
After the ensemble is defined in this way, we proceed to the limit 
in which the boundary shrinks to zero size, recovering the empty static patch as the background to be varied.
The upshot is that the matter Killing energy is seen as 
a (generally) non-thermalized contribution to the (vanishing)
total Brown-York energy of the ensemble, analogous to the contribution
of the energy of matter outside a black hole to the total 
energy of the black hole spacetime, so in the first law
it is {\it subtracted} from the total energy. The temperature 
in the vanishing boundary limit is selected not by a boundary condition, but rather by a saddle point of the boundaryless 
partition function, as explained recently in Ref.~\cite{Banihashemi:2022jys}.


\section{First law for de Sitter with a York boundary} \label{sec:firstlaw}

To attribute a thermodynamical meaning to the GH first law \eqref{GH2},
it must be placed in the context of a well-defined ensemble. Unlike 
the case of a black hole, the static patch of de Sitter space has no 
asymptotic region where the boundary conditions defining the ensemble
can be specified. Thus, to get a thermodynamic handle on the meaning,
we introduce an artificial boundary with Dirichlet boundary conditions inside the static patch, and later take the limit where the boundary is vanishingly small. We call this 
a ``York boundary'', since it was York \cite{York:1986it} who introduced such boundaries in the context of black hole thermodynamics and emphasized that, because of diffeomorphism invariance, gravitational thermodynamics is necessarily anchored at a boundary (see also \cite{Martinez:1996vy} for a general discussion).

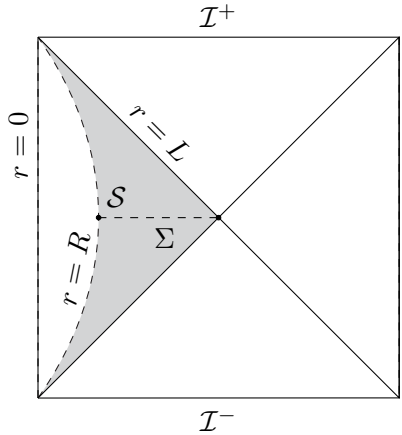
\begin{figure}[t!]
\centering
\begin{tikzpicture}[scale=1.2]
	\pgfmathsetmacro\myunit{4}
	\draw[dashed]	(0,0)			coordinate (a)
		--++(90:\myunit)	coordinate (b);
	\draw (b) --++(0:\myunit)		coordinate (c)
							node[pos=.5, above] {$\mathcal{I}^+$};
	\draw[dashed] (c) --++(-90:\myunit)	coordinate (d);
    \fill[fill=Gray, fill opacity=.4] (a) to[bend right=35]  (b) -- (2,2) -- (a);
    \path (b) to[bend left=35] node[pos=.65,above,sloped] {$r=R$} (a);
    \draw[dashed, Black, name path=rB] (a) to[bend right=35] (b);
    \draw (d) -- (a) 		node[pos=.5, below] {$\mathcal{I}^-$};
    \draw (b) -- (d) node[pos=.3, above, sloped] {$r = L$} -- (c) -- (a);
    \draw (a) -- (b) node[pos=.7, above, sloped] {$r = 0$};
    \draw (c) -- (d);
    \draw[dashed, name path=Sigma] (0.67,2) -- (2,2) 
    node[pos=.55, below] {$\Sigma$};  
    \path[name intersections={of=rB and Sigma, by={int}}] (int) --++(0:1) node[pos=0.2, above] {$\mathcal{S}$};
    \filldraw (int) circle (0.025cm);
    \filldraw (2,2) circle (0.025cm);
\end{tikzpicture}
\caption{\small A York   boundary  (dashed curve) at radius $r=R$ in the dS static patch. The      boundary splits the static patch into two systems: the ``pole patch'', i.e., the white region between the pole and the boundary, and the ``horizon patch'', i.e., the shaded region from the boundary to the   horizon (the terminology comes from \cite{Coleman:2021nor}).   
}
\label{fig:dSbdryB}  
\end{figure}

We assume the boundary is a round 2-sphere   
with fixed area $4\pi R^2$, and that a proper temperature $T$ is fixed at the boundary, and consider the thermodynamic
system between this boundary and the cosmological horizon, called the ``horizon patch'' (see the shaded region in Figure~\ref{fig:dSbdryB}).\footnote{The reservoir needed to physically fix this temperature is not part of the system, and need not be physically modeled here, since in any case we will be taking the limit where the boundary vanishes.   In fact, it is not clear whether Dirichlet boundary conditions on a timelike boundary define a well-posed initial boundary value problem \cite{An:2021fcq}, 
or if  the corresponding system is stable \cite{Andrade:2015gja}. These are important foundational questions for quasi-local gravitational thermodynamics, but we presume they are irrelevant in the vanishing boundary 
limit.
}   
If, for each boundary radius $R$, we
choose the temperature $T$ 
to match the Gibbons-Hawking temperature 
defined with respect to
the Killing vector normalized at the boundary, 
the stationary point of the partition
function will be a Euclidean 
horizon patch of empty dS space, 
and in the limit $R\rightarrow0$ that 
will coincide with the full empty static patch 
with $T=T_{\rm GH}$. The equilibria in this sequence are, 
however, marginally stable. 
As shown in~\cite{Banihashemi:2022jys}, 
and reviewed in Appendix \ref{app:Eucact},
they are local maxima of the
spherical, reduced action; but, if the temperature is 
higher by an arbitrarily small amount, they
become metastable. We shall therefore consider the
limit of a sequence of marginally
stable de Sitter patches,
keeping in mind that to ensure at least 
metastability 
we should in principle 
choose the temperature to be slightly higher
along the sequence. 
Alternatively, 
one could consider instead a sequence of microcanonical 
ensembles, which may be stable.

 In Appendix~\ref{app:Smarrfirstlaw} we show that the on-shell
 variations\footnote{Since we consider a sequence of metastable 
 configurations that lie arbitrarily close to the on-shell marginally stable configuration, the first law for variations
away from an on-shell configuration may be used (cf.\ Appendix \ref{app:Eucact}).}
 of the Brown-York quasi-local energy $E_{\rm BY}$ \cite{Brown:1992br}, matter Killing energy $E_{\rm m}$, 
and horizon entropy $S_{\text{BH}}=A/4G\hbar$ are related by a ``first law'' of the form\footnote{If the area of the round 2-sphere $\mathcal S$ is not kept fixed, then there would be an additional term on the right side of  the first law, $- P \delta A_{\mathcal S}$, where $P$ is the Brown-York surface pressure. We include the boundary area variation in the derivation
in
Appendix~\ref{app:Smarrfirstlaw}, cf.~\eqref{eq:quasifirstlawmattappv2}.}
\beq\label{dEBYH}
\delta E_{\rm BY} = \delta E_{\rm m} + T\,\delta S_{\text{BH}}\, .
\eeq
The Brown-York (BY) energy is defined (up to a possible constant counter-term)
by 
\beq E_{\rm BY}=-\frac{1}{8\pi G}\oint_{\mathcal{S}} d^{2}x\sqrt{\sigma}\,k\,,\label{eq:BYenemain}
\eeq
where the integral is over a spatial cross-section $\mathcal{S}$ of the boundary,
$k$ is the trace of the extrinsic curvature of the boundary as embedded in a spatial slice (defined with respect to normal pointing toward the outside of the system, which here is toward the region with smaller spheres),
and $\sigma$ is the determinant of the spatial metric of the boundary cross section. For the case of a spherical boundary cross-section  we have
\beq E_{\rm BY}=-\frac{kR^2}{2 G} \,.\label{eq:BY2}
\eeq
Further, we define the matter Killing energy   with respect to the static patch Killing vector $\xi$ normalized to unity at the York boundary,
\beq E_{\rm m} = \int_\Sigma T_{\mu\nu}\xi^{\mu}d \Sigma^{\nu}\,,
\label{eq:matterKilling}
\eeq
where  $T_{\mu\nu}$ is the matter stress-tensor, and $\Sigma$ is the spatial slice between the York boundary and the horizon.

Since thermodynamics in the static patch is puzzling, 
it is helpful to first consider how the analogous first law is understood in the case of black holes.
Then, in the limit that the York boundary recedes to spatial infinity, the left-hand side of \eqref{dEBYH} becomes the variation of the total energy of the spacetime, i.e., of the ADM mass, the temperature $T$ becomes the Hawking temperature measured at infinity, and the entropy variation arises from the area variation of the black hole horizon. The matter term then refers to matter outside the black hole, for instance a shell supported by its own stress. Such a shell might or might not be in thermal equilibrium with the ensemble; in either case the first law relation holds.
 
The thermodynamic interpretation of the terms 
in the first law for variations of the static patch
is the same as in the black hole case, {\it mutatis mutandis}.  
But now we can contemplate the limit in which 
the boundary radius $R$ goes to zero. The mean extrinsic curvature
approaches the flat space result $k \rightarrow -2/R$, so 
$E_{\rm BY}\rightarrow 0$ as $R \to 0$. Also, since the Killing vector was 
normalized at the boundary, it becomes normalized at the center
of the static patch, so $T\rightarrow T_{\rm GH}=\hbar/2\pi L$.
Our first law thus reduces in the limit to the GH 
first law \eqref{GH2}. 
 The temperature $T_{\rm GH}$
 arises in this
limit because, for each boundary radius $R$, we have chosen the ensemble
temperature so that the saddle point of the partition function 
would correspond to empty dS. Once the boundary has shrunk to 
zero radius, however, it no longer specifies any ensemble; nevertheless, 
a temperature equal to the proper GH temperature
at the center of the static patch is
determined by the inverse of the circumference of a great circle of the round $S^4$ saddle
of the boundaryless partition function~\cite{Banihashemi:2022jys}.

It is now clear how to understand the minus sign in the GH first
law. The confusion --- for those who, like us, were confused --- arose
due to a mistaken interpretation of the matter Killing energy term 
as referring to energy that is thermalized on account 
of being beyond the horizon. 
But this matter is {\it not} 
beyond the horizon; rather, it is accessible in 
the static patch. 
It is therefore analogous to energy in the region
{\it outside} a black hole, which does not contribute to horizon entropy.
The variation of the horizon entropy is therefore 
the inverse temperature times
the variation of the total (BY) energy (which vanishes as $R\rightarrow0$) 
{\it minus} the contribution of the matter Killing energy.
Part of the variation of the vanishing total energy is 
the matter contribution, which may or may not be thermalized, and 
part is the thermalized contribution captured by the horizon entropy
variation.  


\section{Comments on the first law}

\subsection{Thermalized matter and generalized entropy}
\label{thermat}

If matter is  present in the unperturbed background spacetime, the quasi-local first law \eqref{dEBYH} has an additional term on the right side (see Appendix~\ref{app:Smarrfirstlaw} for a derivation):
  \beq
  \delta E_{\text{BY}} = T\,\delta S_{\text{BH}} + \delta E_{\rm m} +\int_\Sigma dV \! \sqrt{-\xi^2} \frac{1}{2}T^{\mu\nu}\delta g_{\mu\nu}\,. \label{firstlawbackgroundenergy}
  \eeq
If the matter is treated as 
a  perfect fluid  
and is in local equilibrium, as in the classic Bardeen-Carter-Hawking
(BCH) paper \cite{Bardeen:1973gs} (see also \cite{Iyer:1996ky}), 
then the two  matter terms  satisfy a thermodynamic first law for the fluid alone, which 
for a fluid at rest with respect to the Killing vector $\xi$
takes the form: 
\beq \delta E_{\rm m} + \int_\Sigma dV \! \sqrt{-\xi^2} \frac{1}{2}T^{\mu\nu}\delta g_{\mu\nu}=
 \int_\Sigma dV \! \sqrt{-\xi^2} 
 (T_{\rm m}  \delta s_{\rm m}+ \mu\, \delta n)\,,
\eeq
where $T_{\rm m}$ is the local proper temperature and $s_{\rm m}$ is 
the entropy density of the fluid, 
$\mu$ is the local proper chemical potential, and $n$ is
the fluid particle number density. 
If the fluid is in global equilibrium in the ensemble at
the temperature $T$ fixed at the York boundary, then the Tolman 
relation holds, $\sqrt{-\xi^2}\, T_{\rm m} = T$, so that the 
net fluid entropy variation becomes $T\,\delta S_{\rm m}$. 
In this case the matter entropy variation $\delta S_{\rm m}$ 
combines with the horizon entropy variation to form  the total (generalized) entropy variation, such that the quasi-local first law takes the form  
\beq
\delta E_{\text{BY}} = T\,\delta S_{\text{gen}} +
\int_\Sigma dV \! \sqrt{-\xi^2} \mu\, \delta n\,,
\eeq 
where $S_{\text{gen}}\equiv S_{\text m} + S_{\text{BH}}$.
At fixed  BY energy and fluid particle number density, 
the first law becomes
the statement that the 
generalized entropy is stationary, $\delta S_{\text{gen}} |_{E_{\text{BY}},n}=0$.   
 
An important 
special case is when the background state is
empty de Sitter spacetime, and the matter consists
of quantum fields in the  de Sitter   
vacuum state. In semiclassical gravity, where  the stress-energy tensor is replaced by its quantum expectation value but the metric
remains classical,   the ``quantum corrected'' first law of the static patch with a boundary is
\beq
\delta E_{\text{BY}} = \delta \langle E_{\rm m} \rangle + T\,\delta S_{\text{BH}}\,, \label{quantumfirstlaw}
\eeq
where $\langle E_{\rm m} \rangle$ is the  expectation value of the Killing energy.
The  de Sitter vacuum  state of the quantum matter in the static patch is
thermal with respect to the Hamiltonian that generates time translations of the static patch, 
which is equal to the Killing energy~\eqref{eq:matterKilling}  where the stress tensor is now an operator. The expectation value of 
the Killing energy variation  is therefore equal to the Killing 
temperature $T$
times the von Neumann entropy variation,\footnote{This is a case of what
is commonly called ``the first law of entanglement" \cite{Blanco:2013joa} because,
although it is nothing but an example of the Clausius relation for a Gibbs state, the entropy it refers to is 
entanglement entropy if the state variation is restricted 
to pure states.}
\beq
\delta \langle E_{\rm m} \rangle = T\,\delta S_{\text{vN}}\,.
\eeq
In the first law \eqref{quantumfirstlaw} the von Neumann entropy variation combines with  the Bekenstein-Hawking entropy variation to form the generalized entropy variation: $\delta E_{\text{BY}}\! =\! T\,\delta S_{\text{gen}}.$
Hence, at fixed BY energy,   
the quantum first law  for variations away from the 
de Sitter vacuum is the statement that
the generalized entropy is stationary, $\delta S_{\text{gen}}|_{E_{\text{BY}}}=0$. 
That is, the stationarity of the generalized entropy holds in this microcanonical ensemble.\footnote{This has been recently shown as well in the context of Jackiw-Teitelboim gravity in \cite{Pedraza:2021cvx,Svesko:2022txo}.}
If the boundary size goes to zero, the   BY energy vanishes, so the fixed energy requirement can be dropped, which implies the generalized entropy is stationary under {\it all} variations.  
The fact that the generalized entropy is stationary in the 
de Sitter vacuum appears consistent with the idea 
that this 
state describes semiclassical fluctuations around the saddle point of the 
zero energy microcanonical ensemble, i.e., the maximally mixed
state in the Hilbert space of states of a region 
surrounded by a horizon. (For related ideas see 
\cite{Banks:2000fe},\!\!\cite{Fischler},\!\!\cite{Banks:2003ta},\!\!\cite{Banks:2006rx},\!\!\cite{Dong:2018cuv},\!\!\cite{Banihashemi:2022jys},\!\!\cite{Chandrasekaran:2022cip}.)


\subsection{Physical process first law}
 \label{sec:physicalprocess}

\begin{figure}[t!]
\centering
\includegraphics[width=5cm]{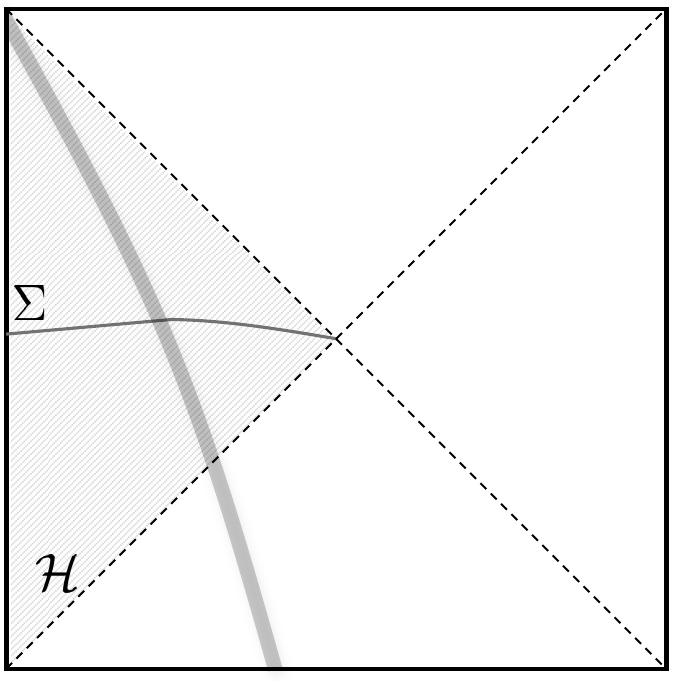}
\caption{\small Matter infalling across the past horizon $\mathcal{H}$ of a dS static patch.
}
\label{fig:infalling} 
\end{figure}

The first law \eqref{dEBYH} discussed in the previous section is a ``stationary state'' version, in which two nearby solutions of the field equations --- namely empty dS and a perturbation of that --- are compared. In that setting, 
we accounted for the minus sign with a thermodynamic interpretation of the law.
In this section we discuss a 
``physical process'' version of the first law, 
which offers a different, dynamical perspective on the 
reason for the minus sign.

The perturbed static patch can arise dynamically
via a (small) flow of matter through the past horizon; see Figure \ref{fig:infalling}. The matter energy momentum tensor is related 
to the Ricci tensor by Einstein's equations, and the null-null component
of the Ricci tensor on the horizon drives a focusing of the 
horizon generators via the Raychaudhuri equation, 
perturbing the horizon area. 
The equation for evolution of the horizon area can be integrated, given 
a boundary condition. In the case of a black hole perturbed by infalling matter \cite{Hawking:1972hy, Carter1979, Wald:1995yp}, 
the boundary condition is that the 
horizon settles down to constant area in the future,
and an essentially identical analysis applies to 
asymptotic Rindler horizons \cite{Jacobson:2003wv,Amsel:2007mh}.
In these cases, one compares the final horizon area to the area 
at the bifurcation surface of the unperturbed background 
spacetime.\footnote{The physical process first law \eqref{eq:physproc} is  valid only for quasistationary processes; in particular, for the higher order terms in the Raychaudhuri equation to be negligible, no caustics should form in the region of integration (i.e.~the perturbed past horizon up to the bifurcation surface in the case of dS). 
In Ref. \cite{Amsel:2007mh} a condition on the size of the infalling matter to avoid caustics was obtained, and it was shown that the physical process first law applies to any bifurcate Killing horizon.} 

The change of dS horizon area can be calculated in essentially
the same manner, using the boundary condition that the 
horizon area is constant (and equal to $4\pi L^2$) 
in the asymptotic past or future
in the perturbed solution. If matter carrying positive energy 
falls across the past horizon of a static patch of dS, as
depicted in Figure \ref{fig:infalling}, the horizon generators are focused,
so that the horizon area at the bifurcation surface of the background solution
is {\it less} than $4\pi L^2$. In this way, addition of 
positive matter energy to the static patch leads to a {\it decrease}
of horizon entropy, hence the minus sign in the physical process
form of the first law, 
\be
\delta E_\text{m}^{\mathcal{H}}=-\frac{\kappa}{8\pi G}\delta A\, .
\label{eq:physproc}
\ee
Here $\kappa$ is the surface gravity\footnote{We take the Killing vector to be future pointing in the static patch and on the horizon. If the surface gravity is defined by one of the standard linear formulas (e.g., $\xi^\mu \nabla_\mu \xi^\nu=\kappa \xi^\nu$) so that it is positive on the future horizon, it is then negative on the past horizon, since the Killing vector shrinks with respect to affine parameter along the horizon Killing flow. However, 
$\kappa$ can also be defined as positive by definition, which is 
what we do in this paper.} and $\delta E_\text{m}^{\mathcal{H}}=\int_{\mathcal{H}} \delta T_{\mu \nu} \xi^\mu d\Sigma^\nu$ is the Killing energy flux across the horizon $\mathcal{H}$, defined with respect to the horizon generating Killing field $\xi^{\mu}$ of the stationary background spacetime. 
To relate the physical process version to the stationary comparison version
of the first law one can use 
the conservation relation, $\nabla^\nu(\delta T_{\mu\nu}\xi^\mu)\!=\!0$, which 
holds to   leading order in the perturbation, and invoke Stokes' theorem to equate the flux of Killing energy across 
$\mathcal{H}$ to the flux across a spatial slice $\Sigma$ of the static patch. 
 
A similar analysis can be done for a time-reversed process, in which a flux of energy exits the static patch through the future horizon. This is analogous to the case of matter falling into a black hole, which produces an increase of the horizon area. The horizon generators must 
start out expanding so that, after the flux of matter across the horizon 
focuses them, they wind up with vanishing expansion in the asymptotic future.
This analogy was invoked in Ref.~\cite{Anninos:2012qw}
to give an information-theoretic explanation for 
the minus sign in the dS first law: information available in the 
static patch about the matter is
lost, so the entropy increases. In effect, the matter energy is thermalized.
Conversely, when matter energy enters through the past horizon
of the patch, the available information increases, so the entropy decreases.

 
 \subsection{Three-dimensional Schwarzschild-de Sitter solution}
\label{sec:3d}

The case of three-dimensional spacetime is 
different from that in higher dimensions, since there are no classical de Sitter black holes in three dimensions.\footnote{Semi-classical backreaction due to conformal fields can induce a black hole horizon, however, leading to a three-dimensional quantum de Sitter black hole \cite{Emparan:2022ijy}.} Rather, three-dimensional SdS is equivalent to a conical defect with a single cosmological horizon \cite{Deser:1983nh}. The defect arises due to a point particle of rest mass $m$ with Killing energy $E_{\text{m}}=-P^{\mu}\xi_{\mu}=m\sqrt{-\xi^{2}}$, where $P^{\mu}=mu^{\mu}$ is the 4-momentum. The point particle may be enclosed inside an arbitrarily small system boundary radius $R$, so the BY energy need not vanish in the limit
$R\to0$. The variation of the BY energy of the ``horizon patch'' due to a variation $\delta m$ of the mass of a point particle  at rest at the center of the ``pole  patch'' (cf.~Figure \ref{fig:dSbdryB}) is equal to the variation of {\it minus}  the Killing energy of the point mass.\footnote{If the Killing vector $\xi$ is normalized to unity at the boundary, then the   matter Killing energy variation of the point mass at rest at the pole is equal to minus the  horizon patch  BY energy variation for nonzero $R$ as well,  and is given by $\delta m$  multiplied by the norm of $\xi$ at the location of the particle.  
In fact, this holds also in higher dimensions.  According to the first law for the  pole patch,  the matter Killing energy variation
is equal to the pole patch BY energy variation, which is minus the horizon patch BY energy variation  since they are defined with respect to opposite normal directions.  In the $R \to 0$ limit  both the BY energy of and   Killing energy variation in the pole patch vanish  in $d\ge4$ spacetime dimensions, but in $d=3$ dimensions with a point particle at the center  
they remain  nonzero.}  This provides another understanding of the origin of the minus sign in the three-dimensional case. Unlike the interpretation offered in Ref. \cite{Spradlin:2001pw}, which was that $-\delta m$ in the first law be viewed as the variation of the Killing energy outside the static patch, here it is seen as the limit of $\delta E_{\rm BY}$ as a boundary enclosing the point mass shrinks to zero size.

More precisely, including a York boundary    with fixed area    in between the cosmological horizon and the origin, the quasi-local first law  for the horizon patch  is  
\beq \delta E_{\text{BY}}=T\,\delta S_{\text{BH}}\, .\eeq
Notice there is no matter Killing energy variation $\delta E_{\text m}$ in this context since the only matter is the point mass
at the center of the static patch, which lies outside the system between the  boundary and cosmological  horizon. Moreover, the BY energy is given    by  $E_{\text{BY}}=\frac{1}{4 G}\sqrt{(1- 4 G m )^2 - R^2/L^2} $ (see Appendix \ref{app:Eucact}). When the York boundary shrinks $R\to0$, we find
\beq \delta(-m)=T_{\text{GH}}^{\text{dS}}\delta S_{\text{BH}}\;,\label{eq:3dfirstlaw}\eeq
where $T_{\text{GH}}^{\text{dS}}= \hbar /2\pi L$. Since the pure de Sitter GH temperature appears here, this relation can also be interpreted as a first law for variations from dS$_3$ to SdS$_3$. Indeed, the   left-hand side   is equal to  minus the Killing energy variation in the static patch
due to the addition of a point mass in empty de Sitter space, since $  \delta E_{\text{m}}= \sqrt{-\xi^2}\delta m = \delta m$, where the norm of $\xi$ is   evaluated at the center of the empty dS$_3$ static patch.  
Hence, the relation \eqref{eq:3dfirstlaw} is nothing but the GH first law \eqref{GH2} applied to a point mass in three-dimensional de Sitter.

It is worth comparing the first law (\ref{eq:3dfirstlaw}) with the first law for  SdS$_3$  proposed by \cite{Klemm:2002ir,Spradlin:2001pw,Balasubramanian:2001nb},
\beq
\delta (- M )= T_{\text{GH}}^{\text{SdS}} \delta S_{\text{BH}}\,,
\eeq
where $M$ is the   mass parameter in the metric, and $T_{\text{GH}}^{\text{SdS}}=\hbar \sqrt{1-8GM}/2\pi L$ is  the GH temperature in the SdS spacetime defined with respect to the Killing vector given by $\xi=\partial_t$ in the coordinate system of \eqref{eq:SdSmet}.  
While the definitions of mass and temperature are different, the two first laws are indeed consistent, since $m$ and $M$ are related via $\sqrt{1-8GM}=1-4Gm$, such that the variations are related by $\delta m =  \delta M / \sqrt{1-8GM}.$ Note that while in \cite{Spradlin:2001pw}  $-M$ was  interpreted as the energy outside the horizon, in our formulation  $-m $ arises from the Brown-York energy as the boundary enclosing the point particle shrinks to vanishing size.

\subsection{York boundary  and de Sitter holography}

Thus far we have focused on the thermal system extending between the York boundary and the cosmological horizon, the shaded region in Figure \ref{fig:dSbdryB}. Consider now the non-shaded region of the static patch between the pole and the York boundary, a.k.a.~the ``pole patch''. There is  a quasi-local first law for the pole patch, given by
\beq \label{polepatchlaw}\delta E_{\text{BY}} = \delta E_{\rm m}-P\delta A_{\mathcal{S}}\;.\eeq
Here we explicitly allowed for variations of the area $A_\mathcal{S}$ of the boundary cross-section. The area is conjugate  to   the ``surface pressure'' $P$, which is proportional to the trace of the spatial stress tensor (see Appendix \ref{app:Smarrfirstlaw}).
Note there is no thermalized entropy variation since the cosmological horizon is hidden behind the York boundary.
In the limit the York boundary approaches the cosmological horizon, the BY energy vanishes, however, the pressure-area variation becomes $-P\delta A_{\mathcal{S}}\to\kappa\delta A/8\pi G$, thereby recovering the Gibbons-Hawking first law (\ref{GH2}) for de Sitter horizons.\footnote{In the horizon limit the BY stress tensor
for the system that includes the horizon
becomes the membrane paradigm stress tensor \cite{Parikh:1997ma}, which has pressure $\kappa/8\pi G$.
The opposite sign appears here because \eqref{polepatchlaw} applies to the pole patch, for which the BY stress tensor is defined with respect to the opposite normal to the York boundary.
}

The York boundary to horizon limit reminds us of a candidate proposal for a holographic description of the dS static patch \cite{Susskind:2021dfc,Susskind:2021esx,Shaghoulian:2021cef}. In this picture the holographic dual theory of a static patch lives at the (stretched) horizon. The maximum entropy associated to the dual theory on the stretched horizon is large enough to describe the entire static patch.  In fact, it is argued the entropy of the cosmological horizon is equal to the entanglement entropy of the pair of   dual quantum theories living on the two stretched horizons.  
There is  another proposal for static patch holography \cite{Anninos:2011af} (see also \cite{Anninos:2017hhn}) where the dual theory lives instead on a worldline near the origin $r=0$.  From the quasi-local perspective, this version of dS holography coincides with the shrinking boundary limit of the system defined between the York boundary and the horizon, the ``horizon patch''. Thus, the two proposals for static patch holography correspond to two limits of the York boundary in de Sitter space. 

The question arises how these two proposals can be   consistent with each other.
In fact, a York boundary of finite radius appears to interpolate between these two proposals of static patch holography, placing them in a single framework, where the boundary offers a fitting location to anchor the dual quantum theory. Further, it has been argued that in de Sitter space the UV/IR connection appears to be inverted compared to standard AdS/CFT \cite{Leuven:2018ejp}, such that long distances in the bulk correspond to low energies in the dual theory. Thus, perhaps moving the boundary   from the pole to the cosmological horizon corresponds to a flow from the UV to IR of the dual theory, respectively \cite{Svesko:2022txo}. Alternately, according to \cite{Coleman:2021nor}  the movement of the York boundary corresponds to a deformation  of a holographic conformal field theory.  The precise nature of the dual quantum theory living on the York boundary in the static patch remains an important open question.


\noindent \noindent\section*{Acknowledgments}
We are grateful to Dionysios Anninos and Erik Verlinde for useful discussions. The research of BB and TJ was supported in part by 
National Science Foundation
grant PHY-2012139. AS is supported by the Simons Foundation via \emph{It from Qubit: Simons Collaboration on quantum fields, gravity, and information}, and EPSRC.  MV is supported by the Republic and Canton of Geneva and the Swiss National Science Foundation, through Project Grants No. 200020-182513 and No. 51NF40-141869 The Mathematics of Physics (SwissMAP).  Some of the work of TJ and MV was conducted at the Peyresq Physics 2022 conference, supported in part by OLAM.


\appendix

\section{Quasi-local Smarr relation and first law} \label{app:Smarrfirstlaw}

Here we derive the quasi-local Smarr relation and first law, including matter, in the presence of a   York boundary. We focus on a York boundary in the static patch of $d$-dimensional de Sitter space, though also extend the formalism to a Schwarzschild-de Sitter black hole. The computations are quite similar to those in \cite{Svesko:2022txo} for Jackiw-Teitelboim gravity in two-dimensional de Sitter space, except that here we consider general relativity plus matter fields in $d$ dimensions.

\subsection*{Geometric set-up}

Consider a $d$-dimensional spacetime $\mathcal{M}$ equipped with  metric $g_{\mu\nu}$. We now perform a $(d-2)+1+1$-dimensional split of the spacetime. We consider a $(d-1)$-dimensional timelike   York boundary~$B$, which has  unit normal    $n_{\mu}$, and induced metric $\gamma_{\mu\nu}=-n_{\mu}n_{\nu}+g_{\mu\nu}$. We denote the extrinsic curvature of the boundary as embedded in $\mathcal M$ as $\mathcal{K}_{\mu\nu}=\gamma^{\alpha}_{\;\mu}\nabla_{\alpha}n_{\nu}=\nabla_{\mu}n_{\nu}$. 
We foliate $\mathcal{M}$ by $(d-1)$-dimensional spacelike hypersurfaces $\Sigma$, with   timelike unit normal $u_{\mu}$ and induced metric $h_{\mu\nu}=u_{\mu}u_{\nu}+g_{\mu\nu}$. The extrinsic curvature of $\Sigma$ is denoted as $K_{\mu\nu}=h^{\alpha}_{\;\mu}\nabla_{\alpha}u_{\nu}=\nabla_{\mu}u_{\nu}$.  We assume the background has a timelike Killing vector field $\xi^{\mu}$ orthogonal to the $\Sigma$ foliation, with norm $\sqrt{-\xi^{2}}=N$, such that $\xi^{\mu}=Nu^{\mu}$.
 Finally, let $\mathcal{S}$ denote the $(d-2)$-dimensional spatial intersection of the York boundary and $\Sigma$. Assuming the foliation $\Sigma$ is orthogonal to $B$, the normals $u_{\mu}$ and $n_{\mu}$ obey $(u\cdot n)|_{B}=(u\cdot n)|_{\mathcal{S}}=0$. This implies both unit normals are also  normal to $\mathcal{S}$, leading to the $(d-2)+1+1$-dimensional decomposition of the spacetime metric
\beq g_{\mu\nu}=-u_{\mu}u_{\nu}+n_{\mu}n_{\nu}+\sigma_{\mu\nu}\;,\label{eq:211split}\eeq
where $\sigma_{\mu\nu}=-n_{\mu}n_{\nu}+h_{\mu\nu}$ is the induced metric on $\mathcal{S}$.  The extrinsic curvature on $\mathcal{S}$ as embedded in $\Sigma$ is denoted by $k_{\mu\nu}=\sigma^{\alpha}_{\;\mu}D_{\alpha}n_{\nu}$, where $D_{\alpha}$ is the covariant derivative intrinsic to $\mathcal{S}$.  
The three extrinsic curvatures $\mathcal{K}_{\mu\nu}$ $K_{\mu\nu}$, and $k_{\mu\nu}$ are related by, cf.~\cite{Brown:1992br},
\beq \mathcal{K}_{\mu\nu}=k_{\mu\nu}-u_{\mu}u_{\nu}n_{\beta}a^{\beta}+(\sigma^{\alpha}_{\;\mu}u_{\nu}+\sigma^{\alpha}_{\;\nu}u_{\mu})n^{\beta}K_{\alpha\beta}\;,\label{eq:extcurvs1}\eeq
where $a^{\beta}=u^{\alpha}\nabla_{\alpha}u^{\beta}$ is the acceleration.
Some useful contractions of this identity are
\beq \mathcal{K}=g^{\mu\nu}\mathcal{K}_{\mu\nu}=k+n_{\alpha}a^{\alpha}\;, \quad u^\mu u^\nu\mathcal{K}_{\mu \nu}=- n_\alpha a^\alpha, \quad  {\sigma_\rho}^\mu u^\nu \mathcal K_{\mu \nu}= - {\sigma_\rho}^\alpha n^\beta K_{\alpha \beta}\,.\label{eq:extcurvs2}\eeq
Further, from the Killing equation it follows that 
\beq  
 u_{\nu}\nabla^{\nu}\xi^{\mu}=\nabla^{\mu}N\;,\quad a^{\mu}=\frac{1}{N}\nabla^{\mu}  N\;, \quad K_{\alpha \beta}=0\;.\label{eq:usefulsplitrels}\eeq
Lastly,  while in the main text we chose the normalization of $\xi$ such that $N =1$ at the York boundary, here we consider an arbitrary normalization and hence retain factors of~$N$.

\subsection*{Quasi-local Smarr relation}

We derive the quasi-local Smarr relation and first law using the covariant phase space formalism of Iyer and Wald \cite{Wald:1993nt,Iyer:1994ys}. Before including matter fields, we first focus on vacuum general relativity characterized by the Einstein-Hilbert Lagrangian $d$-form, including a cosmological constant $\Lambda$,
\beq L=\frac{\epsilon}{16\pi G}(R-2\Lambda)\;,\label{GRL}\eeq
where $\epsilon$ is the volume $d$-form on $\mathcal{M}$. A total variation of $L$ yields 
\beq \delta L=E_{\mu\nu}\delta g^{\mu\nu}+d\theta(g,\delta g)\;\eeq
where the equation of motion $d$-form $E_{\mu\nu}$ is
\beq E_{\mu\nu}  
=\frac{1}{16\pi G}(G_{\mu\nu}+\Lambda g_{\mu\nu})\epsilon\;,\eeq
and the symplectic potential $(d-1)$-form $\theta$ is 
\beq \theta(g,\delta g)=\theta\cdot\epsilon=\theta^{\alpha}\epsilon_{\alpha}\;,\eeq
with
\beq \theta^{\alpha}=\frac{1}{16\pi G}(g^{\mu\alpha}\nabla^{\nu}\delta g_{\mu\nu}-g^{\mu\nu}\nabla^{\alpha}\delta g_{\mu\nu})\;,\label{eq:vacGRthetav1}\eeq
and $\epsilon_{\alpha}\equiv\epsilon_{\alpha\alpha_{2}...\alpha_{d}}$ is the volume form on $\mathcal{M}$ whose first index is displayed while the remaining $(d-1)$ indices are suppressed.

For every diffeomorphism generated by a smooth spacetime vector field $\xi^{\mu}$, there is an associated Noether current $(d-1)$-form $J_{\xi}$, 
\beq J_{\xi}=\theta(g,\mathcal{L}_{\xi}g)-\xi\cdot L\;,\eeq
where $\mathcal{L}_{\xi}$ denotes the Lie derivative along $\xi$. When the equations of motion are satisfied, $E_{\mu\nu}=0$, $J_{\xi}$ is closed, and hence locally it is an exact form  
\beq J_{\xi}=d Q_{\xi}\;,\label{eq:JdQdef}\eeq
where $Q_{\xi}$ is known as the Noether charge $(d-2)$-form.
Explicitly, on the boundary of $\Sigma$, denoted~$\partial\Sigma$, the Noether charge form is 
\beq Q_{\xi}=-\frac{\epsilon_{\partial\Sigma}}{16\pi G}\epsilon_{\mu\nu}\nabla^{\mu}\xi^{\nu}\;,\eeq
where $\epsilon_{\partial\Sigma}$ is the `volume' form on $\partial\Sigma$; on $\mathcal{S}$ we have $\epsilon_{\mu\nu}=(n_{\mu}u_{\nu}-n_{\nu}u_{\mu})$ as its binormal.

The Smarr relation follows from the integral identity of (\ref{eq:JdQdef}) and applying Stokes' theorem
\beq \int_{\Sigma} J_{\xi}=\oint_{\partial\Sigma}Q_{\xi}\;.\label{eq:intidSmarrapp}\eeq 
In the case of the de Sitter static patch, $\partial\Sigma=\mathcal{S}\cup \mathcal{B}$, where $\mathcal{B}$ denotes the bifurcation surface of the cosmological horizon. Since  $\theta(g,\mathcal{L}_{\xi}g)=0$ for Killing vectors, we find for the Lagrangian~\eqref{GRL}
\beq \int_{\Sigma}J_{\xi}=-\frac{2\Lambda V_{\xi}}{(d-2)8\pi G}\;,\quad V_{\xi}\equiv\int_{\Sigma}\xi\cdot \epsilon\;.\label{eq:Jxiapp}\eeq
Here $V_\xi$ denotes the `Killing volume' \cite{Jacobson:2018ahi} (usually denoted by $\Theta$ in the literature).

Meanwhile, the right-hand side of (\ref{eq:intidSmarrapp}) is  
\beq \oint_{\partial\Sigma}Q_{\xi}=\oint_{\mathcal{S}}Q_{\xi}+\oint_{\mathcal{B}}Q_{\xi}\;.\eeq
where the orientation is taken to be toward the pole for the Noether charge integral over~$\mathcal S$  and away from the pole for the integral over $\mathcal B$.
The second term is proportional to the area of bifurcation surface $\mathcal{B}$,
\beq \oint_{\mathcal{B}}Q_{\xi}=-\frac{\kappa}{8\pi G}A_{\mathcal{B}}\;,\label{eq:entapp}\eeq
where we used $\nabla_{\mu}\xi_{\nu}|_{\mathcal{B}}=-\kappa\epsilon_{\mu\nu}$ and $\epsilon_{\mu\nu} \epsilon^{\mu \nu}=-2$. Further, via the metric decomposition (\ref{eq:211split})
\beq Q_{\xi}|_{\mathcal{S}}=\frac{N\epsilon_{\mathcal{S}}}{8\pi G}n^{\mu}a_{\mu}\;,\label{eq:NoetherchargeonS}\eeq
 where we inserted (\ref{eq:extcurvs2}) and (\ref{eq:usefulsplitrels}). By adding zero, we can rewrite this as 
\beq Q_{\xi}|_{\mathcal{S}}=-\frac{d-3}{d-2}\frac{Nk\epsilon_{\mathcal{S}}}{8\pi G}+\frac{N\epsilon_{\mathcal{S}}}{d-2}s^{\mu\nu}\sigma_{\mu\nu}\;,\label{eq:QxiSapp}\eeq
where
\beq s^{\mu\nu}=\frac{1}{8\pi G}[-k^{\mu\nu}+(n_{\alpha}a^{\alpha}+k)\sigma^{\mu\nu}] \label{eq:spatialstressapp}\eeq
is the spatial stress contribution to the Brown-York stress-energy tensor  \cite{Brown:1992br}.

Altogether, substituting (\ref{eq:Jxiapp}), (\ref{eq:entapp}), and (\ref{eq:QxiSapp}) into (\ref{eq:intidSmarrapp}) yields
\beq -\frac{d-3}{d-2}\frac{1}{8\pi G}\oint_{\mathcal{S}}\left ( \epsilon_{\mathcal S} Nk \right)=\frac{\kappa}{8\pi G}A_{\mathcal{B}}-\frac{1}{d-2}\oint_{\mathcal{S}} \left( \epsilon_\mathcal{S}Ns^{\mu\nu}\sigma_{\mu\nu} \right)-\frac{2\Lambda V_{\xi}}{(d-2)8\pi G}\;.\eeq
For a spherically symmetric spacetime, $N$ and $s^{\mu\nu}\sigma_{\mu\nu}$ may be pulled out of the integral leading to the quasi-local Smarr relation for the dS static patch 
\beq (d-3)E_{\text{BY}}=(d-2)\left[\frac{\kappa}{8\pi G N}A_{\mathcal{B}}-PA_{\mathcal{S}}\right]-\frac{2\Lambda V_{\xi}}{8\pi G N}\;.\label{eq:quasilocalSmarrapp}\eeq
Here $A_{\mathcal{S}}\equiv\int_{\mathcal{S}}d^{d-2}x\sqrt{\sigma}$ is the area of the round ($d-2$)-sphere $\mathcal{S}$, $(d-2)P\equiv s^{\mu\nu}\sigma_{\mu\nu}$ (whose thermodynamic meaning is given in Appendix \ref{app:Eucact}) and  $E_{\text{BY}}$ is the Brown-York quasi-local energy  given in (\ref{eq:BYenemain}).
 Explicit expressions for $P$ and $E_{\text{BY}}$ can be found in (\ref{eq:PHdefapp}) and (\ref{eq:EHdefapp}), respectively, for SdS spacetime.

\subsection*{Quasi-local first law}

The first law follows from the integral identity
\cite{Wald:1993nt}
\beq \int_{\Sigma}\omega(g,\delta g,\mathcal{L}_{\xi}g)=\oint_{\partial\Sigma}[\delta Q_{\xi}-\xi\cdot\theta]=\oint_{\mathcal{S}}[\delta Q_{\xi}-\xi\cdot\theta]+\oint_{\mathcal{B}}\delta Q_{\xi}\;,\label{eq:intidfirstlawapp}\eeq
where we used that $\xi|_{\mathcal{B}}=0$. Here $\omega(g,\delta_{1}g,\delta_{2}g)\equiv\delta_{1}\theta(g,\delta_{2}g)-\delta_{2}\theta(g,\delta_{1})$ is the symplectic current $(d-1)$-form.  If a Hamiltonian $H_{\xi}$ corresponding to the evolution of $\xi$ exists in phase space, then by  Hamilton's equations we have
\beq \delta H_{\xi}=\int_{\Sigma}\omega(g,\delta g,\mathcal{L}_{\xi}g)\;.\eeq
When $\xi$ is a Killing vector  $\omega(g,\delta g,\mathcal{L}_{\xi}g)=0$, such that
\beq - \oint_{\mathcal{B}}\delta Q_{\xi}=\oint_{\mathcal{S}}[\delta Q_{\xi}-\xi\cdot \theta]\;.\label{eq:varidfirstlaw}\eeq

\noindent We now explicitly evaluate each side of this integral relation. The quantity on the left-hand side is simply
\beq - \oint_{\mathcal{B}}\delta Q_{\xi}=\frac{\kappa}{8\pi G}\delta A_{\mathcal{B}}\;,\eeq
since $\kappa$ is constant over $\mathcal{B}$. Evaluating the right-hand side  is more involved. First, from  (\ref{eq:NoetherchargeonS}), the variation of the Noether charge at $\mathcal S$ is
\beq 
\delta Q_{\xi}\big|_{\mathcal{S}}=\frac{N\epsilon_{\mathcal{S}}}{8\pi G}\left[(n\cdot a)\frac{1}{2}\sigma^{\mu\nu}\delta\sigma_{\mu\nu}+ (n\cdot a) \frac{\delta N}{N}+\delta(n\cdot a)\right]\;,
\label{eq:deltaQS}\eeq
where we used $\delta\epsilon_{\mathcal{S}}=\epsilon_{\mathcal{S}}\frac{1}{2}\sigma^{\mu\nu}\delta \sigma_{\mu\nu}$. Next, to determine $\xi\cdot\theta|_{\mathcal{S}}$  we first need the pullback of the symplectic potential   to $B$, for which we use a result from \cite{Iyer:1995kg,Harlow:2019yfa} (ignoring the exact contribution $dC$ on the right side, since $\xi \cdot dC = \mathcal L_\xi C -d( \xi \cdot  C)$ vanishes after integrating over $\mathcal S$)
\beq \theta(g,\delta g)|_{B}=-\frac{\epsilon_{B}}{16\pi G}(\mathcal{K}^{\mu\nu}-\gamma^{\mu\nu}\mathcal{K})\delta\gamma_{\mu\nu}+\delta b\;,\eeq
where $b$ is the Gibbons-Hawking-York boundary $(d-1)$-form on $B$
\beq b=-\frac{\epsilon_{B}}{8\pi G}\mathcal{K}\;.\eeq
Then, with a little effort, we find\footnote{To arrive at this expression we made frequent use of  (\ref{eq:extcurvs2}), the decomposition $\gamma_{\mu\nu}=\sigma_{\mu\nu}-u_{\mu}u_{\nu}$, variations $\delta \epsilon_{B}=\epsilon_{B}\frac{1}{2}\gamma^{\mu\nu}\delta\gamma_{\mu\nu}$, $u^{\mu}u^{\nu}\delta\sigma_{\mu\nu}=0$ and $\delta u_{\mu}=u_{\mu}\delta N/N$ (which follows from     $u_{\mu}=-N\partial_{\mu}t$)  and $\xi\cdot\epsilon_{B}|_{\mathcal{S}}=-N\epsilon_{\mathcal{S}}$.}
\beq 
\xi\cdot\theta|_{\mathcal{S}} =- \frac{N\epsilon_{\mathcal{S}}}{8\pi G} \Big[ -  \delta k-(  n \cdot a+k)\frac{1}{2} \sigma^{\mu\nu} \delta\sigma_{\mu\nu} - (n \cdot a) \frac{\delta N}{N}- \delta (n \cdot a )\Big] - N \epsilon_{\mathcal S} \frac{1}{2} s^{\mu \nu} \delta \sigma_{\mu \nu}     \,,
\label{eq:xidotthetaS}  \eeq where $s^{\mu \nu}$ is the Brown-York spatial stress tensor \eqref{eq:spatialstressapp}.
Putting together (\ref{eq:deltaQS}) and (\ref{eq:xidotthetaS}) yields\footnote{Note the right side can   be expressed as $\frac{1}{8\pi G} \left [ N( n \cdot a) \delta \epsilon_{\mathcal S}- N \epsilon_{\mathcal S}(\delta k + \frac{1}{2} k^{\mu \nu} \delta \sigma_{\mu \nu})\right]$. When evaluated at an arbitrary cross section $\mathcal H$ of the horizon, we have $N \to 0$ and $N (n \cdot a) \to \kappa$, such that $\oint_\mathcal{H} [\delta Q_\xi - \xi \cdot \theta ]=\kappa \delta A_{\mathcal H}/8\pi G.$ \label{footnotecross}}
\beq [\delta Q_{\xi}-\xi\cdot\theta]|_{\mathcal{S}}=-\frac{N}{8\pi G}\delta (k\epsilon_{\mathcal{S}})+ N\epsilon_{\mathcal{S}} \frac{1}{2}s^{\mu \nu}\delta\sigma_{\mu\nu}\;.\eeq  
  Returning to the variational identity~(\ref{eq:varidfirstlaw}), we thus find 
\beq -\frac{1}{8\pi G}\oint_{\mathcal{S}} N\delta(k\epsilon_{\mathcal{S}}) =\frac{\kappa}{8\pi G}\delta A_{\mathcal{B}} -\oint_{\mathcal{S}}N \epsilon_{\mathcal S} \frac{1}{2} s^{\mu \nu}\delta \sigma_{\mu \nu}\;.\eeq
  As a final simplification, we assume $\mathcal{S}$ represents a round $(d-2)$-sphere on a spherically symmetric spacetime, such that $k^{\mu\nu}=\frac{1}{ d-2 }k\sigma^{\mu\nu}$ and hence $s^{\mu \nu}= P \sigma^{\mu \nu}$. Moreover, given a spherically symmetric configuration,   $N$ and $P$ are constant over $\mathcal{S}$,  so they may be pulled out of the integral. Thus, after using $\epsilon_{\mathcal{S}}\frac{1}{2}\sigma^{\mu\nu}\delta \sigma_{\mu\nu}=\delta\epsilon_{\mathcal{S}}$, we arrive at the quasi-local (mechanical) first law 
\beq \delta E_{\text{BY}}=\frac{\kappa}{8\pi G N}\delta A_{\mathcal{B}}-P\delta A_{\mathcal{S}}\;.\label{eq:quasifirstlawapp}\eeq
The form of this quasi-local first law is exactly the same as for a Schwarzschild black hole  \cite{York:1986it,Brown:1992br}.
It does not only hold for a spatial slice between the York boundary and the bifurcation surface of the   horizon, but for any slice between $B$ and a  cross section   of the horizon (see footnote \ref{footnotecross}). 

\subsection*{Adding matter}

We   now     minimally couple dynamical matter fields $\psi$ to general relativity, with Lagrangian form $L_{\rm m}(g,\psi,\nabla_{\mu}\psi)\epsilon$. Following Iyer \cite{Iyer:1996ky} (see also \cite{Jacobson:2018ahi}), an arbitrary variation of the total Lagrangian $L=L_{\rm g}+L_{\rm m}$ is
\beq \delta L=E_{\mu\nu}\delta g^{\mu\nu}+E_{\psi}^{\rm m}\delta\psi+d\theta_{\rm g}(g,\delta g)+d\theta_{\rm m}(\phi,\delta\phi)\;.\eeq
Here we denote the collection $\phi=\{g,\psi\}$, and used
\beq \delta L_{\rm m}=E_{\psi}^{\rm m}\delta\psi-\frac{1}{2}\epsilon T_{\mu\nu}\delta g^{\mu\nu}+d\theta_{\rm m}(\phi,\delta\phi)\;,\eeq
with $E_{\psi}^{\rm m}$ being the matter field equations of motion $d$-form, $T_{\mu\nu}$ is the matter stress-energy tensor, and $\theta_{\rm m}$ is the symplectic potential $(d-1)$-form due to the matter fields. Further, 
\beq E_{\mu\nu}=E^{\text{vac}}_{\mu \nu} - \frac{1}{2}T_{\mu \nu} \epsilon=\left[\frac{1}{16\pi G}(G_{\mu\nu}+\Lambda g_{\mu\nu})-\frac{1}{2}T_{\mu\nu}\right]\epsilon\;.\eeq
Next we  decompose  all relevant forms, $\theta,\omega$, and $J_{\xi}$ into gravitational and matter contributions, \emph{e.g.}, $J_{\xi}=J_{\xi}^{\rm g}+J_{\xi}^{\rm m}$. Assuming the background matter field equations, $E^{\rm m}_\psi=0$, the following variational identities can be proven  
\beq \omega_{\rm m}(\phi,\delta\phi,\mathcal{L}_{\xi}\phi)=\delta J_{\xi}^{\rm m}(\phi,\mathcal{L}_{\xi}\phi)+\frac{1}{2}T^{\mu\nu}(\xi\cdot\epsilon)\delta g_{\mu\nu}-d[\xi\cdot\theta_{\rm m}(\phi,\delta\phi)]\;,\label{eq:omegamv1}\eeq
and
\beq \omega_{\rm g}(g,\delta g,\mathcal{L}_{\xi}g)=\delta J^{\rm g}_{\xi}(g,\mathcal{L}_{\xi}g)-(\xi\cdot E^{\mu\nu}_{\text{vac}})\delta g_{\mu\nu}-d[\xi\cdot \theta_{\rm g}(g,\delta g)]\;.\label{eq:omegagv1}\eeq
Importantly, the total $\omega$ is closed on shell. Moreover, $\delta H_{\xi}$ distributes itself into a gravitational Hamiltonian variation $\delta H_{\xi}^{\rm g}$ and a matter Hamiltonian variation $\delta H^{\rm m}_{\xi}$, formally given by 
\beq \delta H^{\rm g}_{\xi}\equiv \int_{\Sigma}\omega_{\rm g}(g,\delta g,\mathcal{L}_{\xi}g)\;,\quad \delta H^{\rm m}_{\xi}\equiv \int_{\Sigma}\omega_{\rm m}(\psi,\delta\psi,\mathcal{L}_{\xi}\psi)\;.\label{Hamsymp}\eeq
Lastly, note that the Noether currents $J^{\rm m}_{\xi}$ and $J^{\rm g}_{\xi}$ are not closed on shell separately; only the total $J_{\xi}$ is. Specifically, assuming $E^{\rm m}_\psi =0$, the $(d-2)$-Noether charge forms $Q_{\xi}^{\rm m}$ and $Q_{\xi}^{\rm g}$ are defined through  
\beq J^{\rm m}_{\xi}=dQ^{\rm m}_{\xi}-T^{\mu\nu}\xi_{\mu}\epsilon_{\nu}\;,\label{eq:Noethercurrmatt}\eeq
and
\beq J^{\rm g}_{\xi} =dQ^{\rm g}_{\xi}+\frac{1}{8\pi G}(G^{\mu\nu}+\Lambda g^{\mu\nu})\xi_{\mu}\epsilon_{\nu}\;.\label{eq:Noethercurrg}\eeq
We now show      that including matter in this way will modify the quasi-local Smarr relation (\ref{eq:quasilocalSmarrapp}) and first law (\ref{eq:quasifirstlawapp}) by an additional integral contribution over the Cauchy slice~$\Sigma$. 

Consider first the Smarr relation, which follows from the 
integral identity (\ref{eq:intidSmarrapp}) applied to the total Noether current and Noether charge. The left-hand side is expressed as the sum of   
\beq \int_{\Sigma}J^{\rm g}_{\xi}=\int_{\Sigma}[\theta_{\rm g}(g,\mathcal{L}_{\xi}g)-\xi\cdot L_{\rm g}]\;,\quad \text{and} \quad \int_{\Sigma}J^{\rm m}_{\xi}=\int_{\Sigma}[\theta_{\rm m}(\phi,\mathcal{L}_{\xi}\phi)-\xi\cdot L_{\rm m}]\;.\eeq
Using $\theta_{\rm g}(g,\mathcal{L}_{\xi}g)=0$ for Killing vectors,  and   the on-shell Einstein-Hilbert Lagrangian \eqref{GRL}
\beq
L^{\rm{on-shell}}_{\rm g}=\frac{\Lambda \epsilon}{(d-2)4\pi G} - \frac{T \epsilon}{d-2}\;,
\eeq
the integral identity (\ref{eq:intidSmarrapp}) becomes
\beq \int_{\Sigma}J^{\rm m}_{\xi}+\int_{\Sigma}\frac{T}{d-2}\xi\cdot\epsilon -\frac{\Lambda V_{\xi}}{(d-2)4\pi G}=\oint_{\partial\Sigma}(Q^{\rm g}_{\xi}+Q^{\rm m}_{\xi})\;,\eeq
where $V_{\xi}$ is the Killing volume \eqref{eq:Jxiapp}.
Subtracting $\int_{\Sigma}dQ_{\xi}^{\rm m}$ from both sides and using (\ref{eq:Noethercurrmatt}) leads to 
\beq -\int_{\Sigma}T^{\mu\nu}\xi_{\mu}\epsilon_{\nu}+\int_{\Sigma}\frac{T}{ d-2 } \xi\cdot\epsilon -\frac{\Lambda V_{\xi}}{(d-2)4\pi G}=\oint_{\partial\Sigma}Q^{\rm g}_{\xi}\;.\label{ultimatesmarr}\eeq
Specifically, for  de Sitter space with   background matter stress-energy we find the quasi-local Smarr relation
\beq (d-3)E_{\text{BY}}=(d-2)\left[\frac{\kappa}{8\pi G N}A_{\mathcal{B}}-PA_{\mathcal{S}}\right]-\frac{2\Lambda V_{\xi}}{8\pi G N}-\frac{ d-2 }{N}\int_{\Sigma}\left(T^{\mu\nu}-\frac{T}{d-2}g^{\mu\nu}\right)\xi_{\mu}\epsilon_{\nu}\;,\label{eq:quasilocalSmarrmattapp}\eeq
which is a generalization of \eqref{eq:quasilocalSmarrapp} to nonvanishing background matter fields. The unfamiliar minus sign in front of the integral disappears if we insert   $\epsilon_\nu |_\Sigma = - u_\nu (u\cdot \epsilon) $.  
Thus the final term contributes positively to the BY energy, if $ \left(T^{\mu\nu}-\frac{T}{d-2}g^{\mu\nu}\right)\xi_\mu u_\nu \ge 0$.   

A similar argument holds for the first law. From the integral identity (\ref{eq:intidfirstlawapp}) we have
\beq \delta H_{\xi}=\oint_{\partial\Sigma}[\delta Q^{\rm g}_{\xi}-\xi\cdot\theta_{\rm g}(g,\delta g)]+\oint_{\partial\Sigma}[\delta Q^{\rm m}_{\xi}-\xi\cdot\theta_{\rm m}(\phi,\delta\phi)]\;,\label{eq:intmedsteplaw}\eeq
where $\delta H_{\xi}=\delta H_{\xi}^{\rm g}+\delta H_{\xi}^{\rm m}$. Presently, $\delta H^{\rm g}_{\xi}=0$ since $\omega_{\rm g}(g,\delta g,\mathcal{L}_{\xi}g)=0$, while from \eqref{eq:omegamv1}, \eqref{Hamsymp} and \eqref{eq:Noethercurrmatt} it follows that  
\beq\delta H_{\xi}^{\rm m}=\int_{\Sigma}\left[\xi\cdot\epsilon\frac{1}{2}T^{\mu\nu}\delta g_{\mu\nu}-\delta(T^{\mu\nu}\xi_{\mu}\epsilon_{\nu})\right]+\oint_{\partial\Sigma}[\delta Q^{\rm m}_{\xi}-\xi\cdot\theta_{\rm m}(\phi,\delta\phi)]\;.\label{matteridentityham}\eeq
Implementing this with (\ref{eq:intmedsteplaw}) yields
\beq \int_{\Sigma}\left[\xi\cdot\epsilon\frac{1}{2}T^{\mu\nu}\delta g_{\mu\nu}-\delta (T^{\mu\nu}\xi_{\mu}\epsilon_{\nu})\right]=\oint_{\partial\Sigma}[\delta Q^{\rm g}_{\xi}-\xi\cdot\theta_{\rm g}(g,\delta g)]\;,\label{varidultimate}\eeq
where we point out the left-hand side is not equal to the matter Hamiltonian variation $\delta H_\xi^{\rm m}$ in the present setup, since the boundary integral on the right side in \eqref{matteridentityham} could be nonzero. In particular,  $\xi \cdot \theta_{\rm}$ integrated over the York boundary is nonvanishing.  
Then, for a dS background with matter we find the quasi-local first law
\beq \delta E_{\text{BY}}=\frac{\kappa}{8\pi G N}\delta A_{\mathcal{B}}-P\delta A_{\mathcal{S}} -\frac{1}{N}\int_\Sigma \delta \left (T^{\mu\nu}\xi_{\mu}\epsilon_{\nu}\right) +\frac{1}{N} \int_{\Sigma} \xi \cdot \epsilon \frac{1}{2} T^{\mu \nu} \delta g_{\mu \nu} \;.\label{eq:quasifirstlawmattappv2}\eeq
 When the background spacetime is pure de Sitter, and the stress tensor is associated to matter fields and not to the cosmological constant, the second integral on the right   vanishes.\footnote{The derivation of the first law  does not assume that the background matter fields share the Killing symmetry. Therefore, a fluid description that uses potentials which do  not share the same symmetry as the stress tensor   is also covered by this computation. For example, the    cosmological constant can be treated as a perfect fluid, and could contribute to $T^{\mu\nu}$ in \eqref{eq:quasifirstlawmattappv2}. In that case, the second integral does not vanish, but on a maximally symmetric background it combines with the first integral to form $-\int_\Sigma \delta {T_a}^b  \xi^a \epsilon_b $, which is equal to $V_\xi \delta \Lambda / 8 \pi G$ \cite{Jacobson:2018ahi}.} 
Further, we recognize the first   integral as the variation of the Killing energy~\eqref{eq:matterKilling}, $N^{-1}\delta E_{\rm m}$.\footnote{The notation in this appendix is related to that in the main text by: $\epsilon_\nu |_\Sigma = - d \Sigma_\nu = - u_\nu dV.$} Further, note that if we choose the normalization of the Killing vector $\xi$ such that it has unit norm at the system boundary, $N |_B=1$, and fix the area~$A_{\mathcal S}$, we recover the form of the first law \eqref{firstlawbackgroundenergy}  expressed in the main text, which reduces to \eqref{dEBYH} if the background stress tensor vanishes. Lastly, in the limit the boundary shrinks to zero size, then $P\to0$ and $E_{\text{BY}}\to 0$,   thus the quasi-local first law reduces to the GH first law~\eqref{GH2}.

\subsection*{Schwarzschild-de Sitter black hole}

\begin{figure}[t!]
\centering
\begin{tikzpicture}[scale=1.2]
	\pgfmathsetmacro\myunit{4}
	\draw (b) --++(0:\myunit)		coordinate (c)
							node[pos=.5, above] {$\mathcal{I}^+$};
    \fill[fill=Gray, fill opacity=.4] (a) to[bend left=35] (b) -- (2,2) -- (a);
    \draw (a) -- (-2,2) -- (b) node[pos=.5, above, sloped] {$r = r_b$};
    \draw[dashed, Black, name path=rB] (a) to[bend left=35] (b);
    \path (a) to[bend left=35] node[pos=.65,above,sloped] {$r=R$} (b);
	\draw (d) -- (a) 		node[pos=.5, below] {$\mathcal{I}^-$};
	\draw (b) -- (d) node[pos=.3, above, sloped] {$r = r_{c}$} -- (6,2) -- (c) -- (a);
    \draw[decorate, decoration={snake, amplitude=0.5mm, segment length=2.5mm}] (a) -- (-2,0) coordinate (e);
    \draw (e) -- (-2,4) coordinate (f);
    \draw[decorate, decoration={snake, amplitude=0.5mm, segment length=2.5mm}] (f) -- (b); 
    \draw[decorate, decoration={snake, amplitude=0.5mm, segment length=2.5mm}] (c) -- (6,4) coordinate (g); 
    \draw (g) -- (6,0) coordinate (h);
    \draw[decorate, decoration={snake, amplitude=0.5mm, segment length=2.5mm}] (h) -- (d);
    \draw[dashed, name path=Sigma] (-.7,2) -- (2,2) node[pos=.4, below] {$\Sigma$} node[pos=-.45, below] {$\mathcal{B}_{b}$}  node[pos=1, below] {$\mathcal{B}_{c}$};
    \path[name intersections={of=rB and Sigma, by={int}}] (int) --++(0:1) node[pos=0.1, below] {$\mathcal{S}$};
    \filldraw (int) circle (0.025cm);
    \filldraw (2,2) circle (0.025cm);
    \filldraw (-2,2) circle (0.025cm);
\end{tikzpicture}
\caption{\small The Penrose diagram of Schwarzschild-de Sitter spacetime. The   York boundary    lies at $r=R$ between the black hole and cosmological horizon, and hence defines two distinct thermal systems:
the white region between the black hole horizon and the boundary, and the shaded region between the boundary and the cosmological horizon.  
}
\label{fig:SdSbdryB}  
\end{figure}
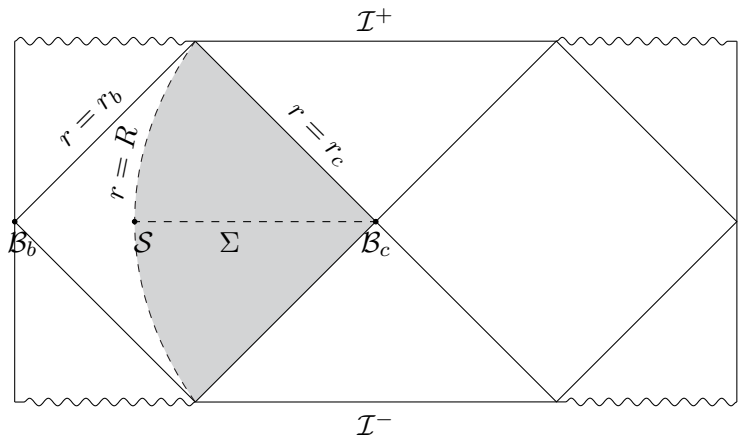

The above derivations all easily extend to the case of a York boundary inside the static patch of a Schwarzschild-de Sitter (SdS) black hole. The essential new feature is that the static patch now has    a black hole horizon $r=r_{b}$, in addition to the cosmological horizon $r=r_{c}$, replacing the pole at $r=0$. Placing the York boundary  between the black hole and cosmological horizons bisects the static patch into two distinct systems: (i) the black hole system  $r\in[r_{b},R]$, and (ii) the cosmological system $r\in[R,r_{c}]$. Then, $\Sigma$ denotes a spatial slice extending between $\mathcal{S}$ and  the bifurcate surface $\mathcal{B}_{b,c}$ of either the black hole or cosmological   horizon  (see Figure \ref{fig:SdSbdryB}).

With this set-up in place, it is straightforward to redo the analysis for either the cosmological or black hole systems separately. For the cosmological system the quasi-local Smarr relation (\ref{eq:quasilocalSmarrapp})   of SdS  becomes 
\beq (d-3)E_{\text{BY},c}=(d-2)\left[\frac{\kappa}{8\pi G N}A_{\mathcal{B}_{c}}-P_{c}A_{\mathcal{S}}\right]-\frac{2\Lambda}{8\pi G N}V^{c}_{\xi}\;. 
\label{quasisdsnew}\eeq
Here $\kappa_{c}$ refers to the   surface gravity of the cosmological horizon,  
  $P_{c}$ is the surface pressure and $V^{c}_{\xi}$ refers to the Killing volume of the cosmological system. 
Likewise, the first law (\ref{eq:quasifirstlawmattappv2}) for SdS is (setting $T^{\mu \nu}=0$ in the background) 
\beq \delta E_{\text{BY},c}=\frac{\kappa_{c}}{8\pi G N}\delta A_{\mathcal{B}_{c}}-P_{c}\delta A_{\mathcal{S}}-\frac{1}{N}\int_{\Sigma}\delta T_{\mu\nu}\xi^{\mu}\epsilon^{\nu}\;.\label{eq:quasifirstlawmattSdSapp}\eeq
Identical expressions hold for the black hole system, where one replaces $\kappa_{c}\to\kappa_{b}$, $\mathcal{B}_{c}\to\mathcal{B}_{b}$, etc. 

It is worth pointing out the Gibbons-Hawking first law of event horizons (\ref{GHfirstlaw}) may be recovered from the above quasi-local first law. In the limit where the York boundary approaches the black hole horizon, $R\to r_{b}$,  the extrinsic curvature trace approaches zero $k\to0$, while $N(n^{\alpha}a_{\alpha})\to-\kappa_{b}$, hence $NP_{c}\to-\kappa_{b}/8\pi G$. Additionally, in this limit $E_{\text{BY},c}\to0$ while $A_{\mathcal{S}}\to A_{\mathcal{B}_{b}}$, and thence, 
\beq \delta E_{\rm m}=-\kappa_{c}\delta A_{c}/8\pi G-\kappa_{b}\delta A_{b}/8\pi G\;.\label{SdSfirstlaw} \eeq
Similarly, in the limit $R \to r_b$ the quasi-local Smarr relation \eqref{quasisdsnew} reduces to the (generalized) Smarr formula for SdS, see e.g. \cite{Dolan:2013ft},
\beq
0= \frac{\kappa_c A_c}{8 \pi G}+\frac{\kappa_b A_b}{8 \pi G} - \frac{2}{d-2}\frac{\Lambda V_{\xi}}{8 \pi G}\,.
\eeq
An identical analysis holds for the black hole system,  where instead the York boundary approaches the cosmological horizon,  $R\to r_{c}$, and $NP_{b}\to-\kappa_{c}/8\pi G$ and so forth. 
Although the first law and Smarr formula for  two event horizons can be recovered in this way from a limit of the first law with a York boundary,   a temperature has not been fixed at the York boundary, so these are not thermodynamic equilibrium relations. In fact, since the two horizons have different surface gravity, the system is not in equilibrium, so the first law in this context is related to a limit of nonequilibrium thermodynamics in which the time dependence is neglected.


\section{Total attractive Killing energy vanishes on a closed slice}
\label{app:zeroKillingenergy}
Here we show that the total matter Killing energy variation  vanishes 
on a closed slice of global de Sitter space, which was stated without proof in Ref.~\cite{Spradlin:2001pw}. In fact, we derive two more general statements about the total matter energy (variation) on a closed spatial slice $\Sigma$ of   stationary spacetimes with a   Killing field $\xi$  and with background matter stress-energy tensor~$T^{\mu \nu}$.  The first statement is about the total (attractive) Killing energy on $\Sigma$ in the background spacetime, and the second one involves metric and matter field variations away from the background spacetime. The results apply in particular to a Cauchy slice $\Sigma$ of a global asymptotically de Sitter space  with matter stress-energy.

First, we remind the reader of the on-shell identity \eqref{ultimatesmarr}.\footnote{  This identity was derived in \cite{Bardeen:1973gs} by
integrating the Killing identity relating two derivatives of the Killing vector to the Ricci tensor, $\nabla_\nu \nabla^\mu \xi^\nu= {R^{\mu}}_\nu \xi^\nu$. The term involving the derivatives of the Killing vector turns into a surface integral  
(which is equivalent to the Noether charge integral in \eqref{ultimatesmarr}) and vanishes on a closed slice. The Ricci tensor term yields the attractive Killing energy density via the Einstein equation.} Since the spatial slice $\Sigma$ is closed,  the boundary integral on the right-hand side of this identity vanishes. Therefore, the following integral over $\Sigma$ is zero 
 \beq
\int_\Sigma \left ( T^{\mu \nu} - \frac{T}{d-2} g^{\mu \nu} + \frac{\Lambda}{(d-2)4\pi G} g^{\mu \nu} \right) \xi_\mu   \epsilon_\nu =0\,.
\label{eq:attractiveK}
\eeq
Note the integrand is not 
the same as the matter Killing energy density \eqref{eq:matterKilling}. In fact,  the Einstein equation implies that 
the quantity between brackets is simply the Ricci tensor $R^{\mu \nu}$ divided by $8 \pi G$. For static spacetimes with $\xi^\mu = N u^\mu$, where $u^\mu$ is the   unit normal to $\Sigma,$  the integrand is proportional to the scalar  $R^{\mu \nu} u_\mu u_\nu$, which via the Raychaudhuri equation governs the focusing of a timelike geodesic congruence initially tangent to $u^\mu$. 
The integrand could thus be called the ``attractive Killing energy density'' or ``focusing Killing energy density''.

 Second, we invoke the variational on-shell identity \eqref{varidultimate}. On a closed slice $\Sigma$ the boundary integral on the right  again vanishes. Thus, we find\footnote{This variational identity can also be obtained from writing the stress tensor trace and the cosmological constant term in \eqref{eq:attractiveK} in terms of the Ricci scalar and then varying the whole integral. The variation of the Ricci scalar   produces minus the first term in \eqref{eq:varattractiveK}, plus a boundary term  which vanishes on a closed slice (see (29) in \cite{Bardeen:1973gs}), while the variation of the Killing energy yields minus the second term. 
 }  
\beq
 \int_{\Sigma}\left [\xi \cdot \epsilon \frac{1}{2}T^{\mu \nu} \delta g_{\mu \nu}-\delta \left (T^{\mu\nu}\xi_{\mu}\epsilon_{\nu}\right) \right]=0\,.
 \label{eq:varattractiveK}
 \eeq
 In empty de Sitter space the first term vanishes, and we are left with the desired result that the first-order variation of the total matter Killing energy   vanishes on  a Cauchy surface, $\delta E_{\rm m}=0$ on~$\Sigma$. As an example, Ref. \cite{Spradlin:2001pw} considered empty three-dimensional de Sitter space and variations towards  dS space  with   two point particles at the poles  (see section \ref{sec:3d}). Since the matter Killing energy variation in the static patch with future pointing Killing vector is equal to the point mass~$m$, whereas in the opposite patch it is $- m  $ because the Killing vector is past pointing, hence the total Killing energy variation vanishes on a closed slice.


\section{On-shell Euclidean action of Schwarzschild-de Sitter} \label{app:Eucact}

Although not directly relevant to this paper, we include here results about the thermodynamics of Schwarzschild-de Sitter (SdS) ensembles with a York boundary that could be useful for future applications. Following York's original paper  \cite{York:1986it} we compute the on-shell Euclidean action for SdS without fluid matter, using static coordinates, and we derive explicit formulae for the thermodynamic quantities discussed in the main text. In \cite{Banihashemi:2022jys} the on-shell Euclidean action for SdS was also computed in a coordinate independent way; our results are in agreement, but here in addition we compute the surface pressure and heat capacity. It is straightforward to extend the analysis below to include fluid matter, as shown in \cite{Hayward:1990zm}.

In static patch coordinates the SdS line element is
\beq ds^{2}=-f(r)dt^{2}+f^{-1}(r)dr^{2}+r^{2}d\Omega_{d-2}^{2}\;,\quad f(r)=1-\frac{r^{2}}{L^{2}}-\frac{16\pi G Mr}{(d-2)A(r)}\;,\label{eq:SdSmet}\eeq
where  $A(r)=\Omega_{d-2}r^{d-2}$ and  $M$ denotes the mass of the black hole. When $0<M<M_{N}$, the blackening factor $f(r)$ has two distinct positive roots: (i) the black hole horizon $r=r_b$ and (ii) the cosmological horizon $r=r_{c}$, where $r_b<r_{c}$.  The Nariai solution $M=M_N$ corresponds to the case where the two horizon radii coincide. 
Each horizon has its own surface gravity $\kappa_{b,c}$ which depends on the normalization of the horizon generating Killing vector. For concreteness let us define the surface gravity  with respect to the time-translation Killing vector~$\xi=\partial_t$, in which case it is equal to 
\beq \kappa_{b,c}=\pm\frac{(d-3)-(d-1)r^{2}_{b,c}/L^{2}}{2r_{b,c}}=\pm\left(\frac{1}{2}\frac{d-3}{d-2}\frac{16\pi G M}{A_{b,c}}-\frac{r_{b,c}}{L^{2}}\right)\;,\label{eq:kappashc}\eeq
where the `$+$' sign refers to the black hole horizon while the `$-$' sign is associated with the cosmological horizon.

Upon Wick rotating the time coordinate $t\to -i\tau$, the Euclidean SdS spacetime has a conical singularity at each horizon $r_{b,c}$. Nonetheless, the Euclidean section for either the black hole or cosmological system (see Figure \ref{fig:SdSbdryB}) can be made smooth by removing the conical singularity at  $r_b$ or $r_{c}$, respectively. The conical singularity at either horizon is smoothed out by making the Euclidean time~$\tau$ is periodic in the inverse Gibbons-Hawking temperature $\beta_{b,c}^{\text{GH}}=\frac{2\pi}{\hbar\kappa_{b,c}}$. The proper length of the boundary at $r=R$ is equal to the inverse Tolman temperature
\beq\label{invTolman}
\beta_{b,c}=\int_{0}^{\beta_{b,c}^{\text{GH}}}\sqrt{f(R)}d\tau=T^{-1}_{b,c}\;.\eeq
We now express the canonical partition function $Z(\beta_{b,c})$ of either system as a gravitational Euclidean path integral, which, in a saddle-point approximation is
\beq Z(\beta_{b,c})\approx e^{-I^{E}_{b,c}}\;,\eeq
where $I^{E}_{b,c}$ is the on-shell Euclidean action. Off-shell, the total Euclidean action is
\beq I=-\frac{1}{16\pi G}\int_{\mathcal{M}_{E}}d^{d}x\sqrt{g}(R-2\Lambda)-\frac{1}{8\pi G}\int_{\partial \mathcal{M}_{E}}d^{d-1}x\sqrt{\gamma}\mathcal{K}\;.\eeq
With respect to the Euclideanized SdS solution, the total on-shell action $I_{b,c}$ for either system is
\beq I^{E}_{b,c}=\mp\frac{\beta_{b,c}}{8\pi G}\frac{ d-2 }{R}A_{\mathcal{S}}\sqrt{1-\frac{R^{2}}{L^{2}}-\frac{16\pi G MR}{(d-2)A_{\mathcal{S}}}}-\frac{A_{b,c}}{4G}\;,\label{eq:onshellacttot}\eeq
where the `$-$' sign on the first term corresponds to the black hole system while the `$+$' is associated with the cosmological system. The limit of \eqref{eq:onshellacttot} where the York boundary coincides with the other horizon, $R \to r_{b,c}$, is ill-defined   since there is a conical singularity at that location in Euclidean SdS. However, by   taking the contribution from the conical singularity properly into account the on-shell    action of the total Euclidean SdS manifold can be shown to be finite  and   equal to 
$I_{\text{tot}}^{\rm E} = - \frac{A_b}{4 G} - \frac{A_c}{4G}$ \cite{Chao:1997osu,Bousso:1998na,Gregory:2013hja,Draper:2022xzl,Morvan:2022ybp}.

From the on-shell action (\ref{eq:onshellacttot})  we may directly compute the thermodynamic energy  
\beq E_{b,c}=\left(\frac{\partial I_{b,c}^{E}}{\partial\beta_{b,c}}\right)_{\hspace{-2mm}A_{\mathcal{S}}}=\pm\frac{1}{8\pi G}\frac{ d-2 }{R}A(R)\sqrt{f(R)}\;,\label{eq:EHdefapp}\eeq
where the positive sign corresponds to the cosmological system and the negative sign for the black hole system. Note the energy $E_{b,c}$ is   equivalent to the quasi-local energy given in (\ref{eq:BYenemain}). Similarly, the thermodynamic entropy   is
\beq S_{b,c}=\beta_{b,c}\left(\frac{\partial I_{b,c}^{E}}{\partial \beta_{b,c}}\right)_{\hspace{-2mm}A_{\mathcal{S}}}-I_{b,c}^{E}=\frac{A_{b,c}}{4G\hbar}\;.\eeq
Therefore, the free energy satisfies $\beta_{b,c}F_{b,c}=I_{b,c}^{E}=E_{b,c}-T_{b,c}S_{b,c}$. As an aside, may include a standard background subtraction contribution to the action $\frac{1}{8\pi G}\int_{B}d^{d-1}x\sqrt{\gamma}\mathcal{K}_{0}$, where $\mathcal{K}_{0}$ is the mean extrinsic curvature at $r=R$ in a pure dS background. Doing so leads to background subtracted energies $\bar{E}_{b,c}$ and $\bar{F}_{b,c}$.  

Further, by fixing the entropy $S_{b,c}$, we define a surface pressure $P_{b,c}$ to be
\beq
\begin{split}
P_{b,c}&=-\left(\frac{\partial E_{b,c}}{\partial A_{\mathcal{S}}}\right)_{\hspace{-1mm}S_{b,c}}=\pm\frac{1}{8\pi G}\left[\frac{d-3}{r}\sqrt{f(r)}+\partial_{r}\sqrt{f(r)}\right]\biggr|_{r=R}\\
&=\pm\frac{1}{8\pi G}\left[\frac{d-3}{R}\sqrt{1-\frac{R^{2}}{L^{2}}-\frac{16\pi GMR}{(d-2)A(R)}}+\frac{\frac{d-3}{d-2}\frac{8\pi GM}{A(R)}-\frac{R^{2}}{L^{2}}}{\sqrt{1-\frac{R^{2}}{L^{2}}-\frac{16\pi GMR}{(d-2)A(R)}}}\right]\;,
\end{split}
\label{eq:PHdefapp}\eeq
where the `$+$' and `$-$' signs correspond to the black hole and cosmological systems, respectively. Nicely, the pressure $P_{b,c}$ coincides with the trace of the spatial stress $s^{\mu\nu}$ (\ref{eq:spatialstressapp}), $P_{b,c}=\frac{1}{d-2}s^{\mu\nu}\sigma_{\mu\nu}$. 

With this on-shell Euclidean analysis in place, we recognize the mechanical first law (\ref{eq:quasifirstlawapp}) as a quasi-local first law of thermodynamics. Indeed, with the thermodynamic quantities above, the Smarr relation (\ref{eq:quasilocalSmarrapp}) and first law (\ref{eq:quasifirstlawapp}) may be verified explicitly.\footnote{\label{foot:dEdM}For instance, at fixed radius $R$ one can check that: $\delta E_{b,c} = \pm \frac{1}{N(R)} \delta M =    \frac{1}{N(R)} \frac{\kappa_{b,c}}{8\pi G  } \delta A_{b,c} $.}

\subsection*{Special case: three-dimensional Schwarzschild-de Sitter}

When $d=3$, the blackening factor of the SdS solution (\ref{eq:SdSmet}) simplifies to  
\beq f(r)=1-8GM-\frac{r^{2}}{L^{2}}\;,
\eeq
for $0<M< 1/8G$. There is now only a   cosmological horizon at $r_{c}=L\sqrt{1-8GM}$, but not a black hole horizon, and the SdS solution is equivalent to a conical defect \cite{Deser:1983nh}.  In fact, in the global extension of $\text{SdS}_{3}$ there are two conical singularities, one at each pole of a two-sphere. The defect arises due to a point mass $m$ at the origin $r=0$, characterized by a stress-energy tensor 
\beq \sqrt{g^{(2)}}T_{\mu\nu}u^{\mu}u^{\nu}=m\delta^{(2)}(r)\;,\eeq
where $g^{(2)}$ refers to the determinant of the metric on a constant $t$ surface $\Sigma$ and $u^{\mu}=f^{-1/2}\partial_{t}^{\mu}$ is the unit normal to   $\Sigma$. The point mass $m$ is   related to $M$ via $4Gm =1-\sqrt{1-8GM}$ \cite{Klemm:2002ir}. 

For the cosmological system between a York boundary at $r=R$ and the     cosmological horizon, the on-shell Brown-York energy, horizon entropy, Tolman temperature, and surface pressure 
are given as a function of $M$ and $R$ by (see also \cite{Coleman:2021nor})
\beq
\begin{aligned}
 E_{\text{BY}}&=\frac{1}{4G}\sqrt{1-8GM-\frac{R^{2}}{L^{2}}}\;,\qquad \qquad   S_{\text{BH}}=\frac{2\pi L\sqrt{1-8GM}}{4G\hbar}\;,   \\ T&=\frac{\hbar\sqrt{1-8GM}}{2\pi L}\frac{1}{\sqrt{1-8GM-\frac{R^{2}}{L^{2}}}}\;,\,\, \,\,\,P=\frac{1}{8\pi G}\frac{R}{L^{2}}\frac{1}{\sqrt{1-8GM-\frac{R^{2}}{L^{2}}}}\;.
\end{aligned}
\eeq
These thermodynamic quantities satisfy the quasi-local first law   
\beq \delta E_{\text{BY}}=T\,\delta S_{\text{BH}}-P\delta A_{\mathcal{S}}\;,\eeq
with $A_{\mathcal{S}}=2 \pi R.$ Since there is no black hole horizon, the limit $R\to0$ sees a vanishing pressure, however, a non-vanishing BY energy, as described in the main text.

\subsection*{Heat capacity and thermal stability}

Of interest is whether thermal systems with a horizon are stable under thermal fluctuations. Indeed, this was a central motivation which led York \cite{York:1986it} to introduce a fictitious boundary in the first place: the heat capacity of an asymptotically flat Schwarzschild black hole system becomes positive  for  a boundary radius  $  R < 3 GM $ in $d=4$, where $M$ is the mean value of the black hole mass in the ensemble at fixed temperature.\footnote{In higher dimensions the heat capacity is positive when $ R< r_b \left (\frac{d-1}{2 (d-3)} \right)^{1/(d-3)}$, where  $r_b$ is the Schwarzschild radius.} The positive heat capacity indicates that the system  is thermally stable. Likewise, we may compute the heat capacity for the black hole and cosmological systems in the SdS solution. At fixed area $A_{\mathcal{S}}$ of the York boundary, the heat capacity $C_{A_{\mathcal{S}}}$ is 
\beq C_{A_{\mathcal{S}}}\equiv T\left(\frac{\partial S}{\partial T}\right)_{\hspace{-1mm}A_{\mathcal{S}}}=\left(\frac{\partial E}{\partial M}\right)_{\hspace{-1mm}A_{\mathcal{S}}}\left(\frac{\partial M}{\partial T}\right)_{\hspace{-1mm}A_{\mathcal{S}}}=\pm\frac{1}{N(R)}\left(\frac{\partial T}{\partial M}\right)^{-1}_{\hspace{-1mm}A_{\mathcal{S}}}\;,
\eeq
where the `$+$' sign refers to the black hole system while the `$-$' sign is associated with the cosmological system. Explicitly, for the black hole system we have
 \beq C_{b,A_{\mathcal{S}}}=\frac{2\pi M}{\kappa_{b}}\frac{\left[1-\frac{R^{2}}{L^{2}}-\frac{r_{b}^{d-3}}{R^{d-3}}\left(1-\frac{r_{b}^{2}}{L^{2}}\right)\right]}{\frac{1}{2}\frac{r_{b}^{d-3}}{R^{d-3}}\left(1-\frac{r_{b}^{2}}{L^{2}}\right)-\left[1-\frac{R^{2}}{L^{2}}-\frac{r_{b}^{d-3}}{R^{d-3}}\left(1-\frac{r_{b}^{2}}{L^{2}}\right)\right] \frac{d-3+2r_{b}^{2}/L^{2}-(d-1)r^{4}_{b}/L^{4}}{(d-3-(d-1)r^{2}_{b}/L^{2})^{2}} }.\eeq
\begin{figure}[t]
\begin{center}
\includegraphics[width=5.7cm]{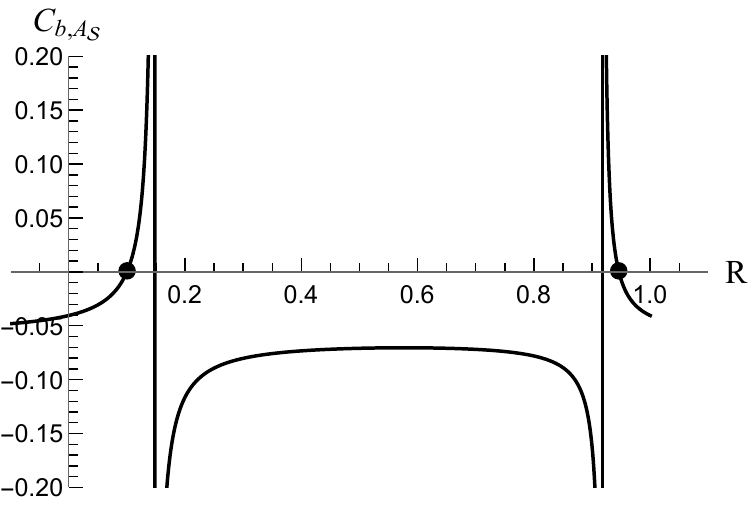}$\qquad\quad$\includegraphics[width=6.7cm]{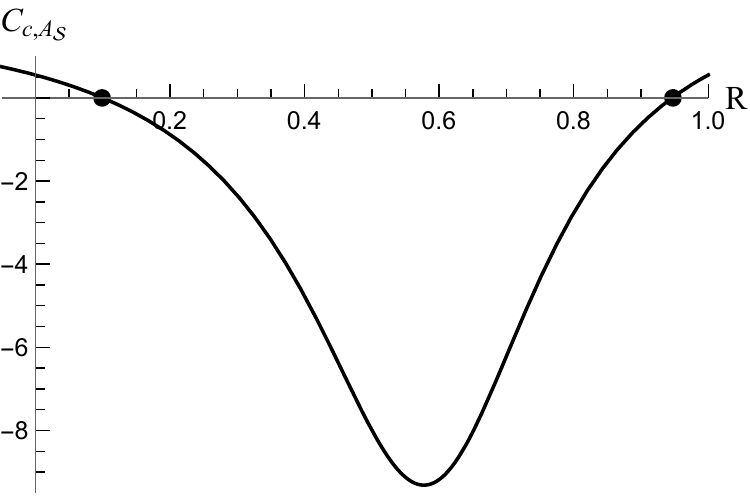}
\end{center}
\caption{\small Heat capacity of black hole system (left) and cosmological system (right) as a function of the boundary radius $R$ for $d=4$. The graphs are meaningful only between the horizon radii (thick circles), where both  heat capacities vanish.  The minimum on the right lies at the Nariai radius $L/\sqrt{3}.$ Here $r_{b}=0.1$ and $L=1$.}
\label{fig:heatcaps} 
\end{figure}
A similar relation holds for the cosmological system, up to an overall minus sign and the replacements $r_{b}\to r_{c}$ and $\kappa_{b}\to\kappa_{c}$.  In Figure \ref{fig:heatcaps} we plot the heat capacities for each system as a function of radius $R$  at fixed mass $M$ and de Sitter radius $L$. In either system, the heat capacities vanish when $R=r_{b,c}$. Moreover,  for the black hole system, we observe $C_{A_{\mathcal{S}}}$ is positive when $r_{b}<R<R_{1}$ and $R_{2}<R<r_{c}$, and negative between $R_{1}<R<R_{2}$, where $R_{1,2}$ refer to two (complicated) values where the heat capacity encounters  discontinuities. Thus, the black hole system is thermodynamically stable for small systems  $R<R_{1}$ and large systems $R>R_{2}$. 
 The heat capacity of the cosmological system, computed from the on-shell action, is negative everywhere between $r_{b}<R<r_{c}$, indicating that the cosmological system in the on-shell configuration is thermodynamically unstable for all values of $R$. 
Note that these stability properties refer to an ensemble at a fixed boundary temperature, which in the saddle point configuration coincides with the Tolman temperature evaluated at the boundary radius.

\begin{figure}[t!]
    \centering  \includegraphics[scale=0.2]{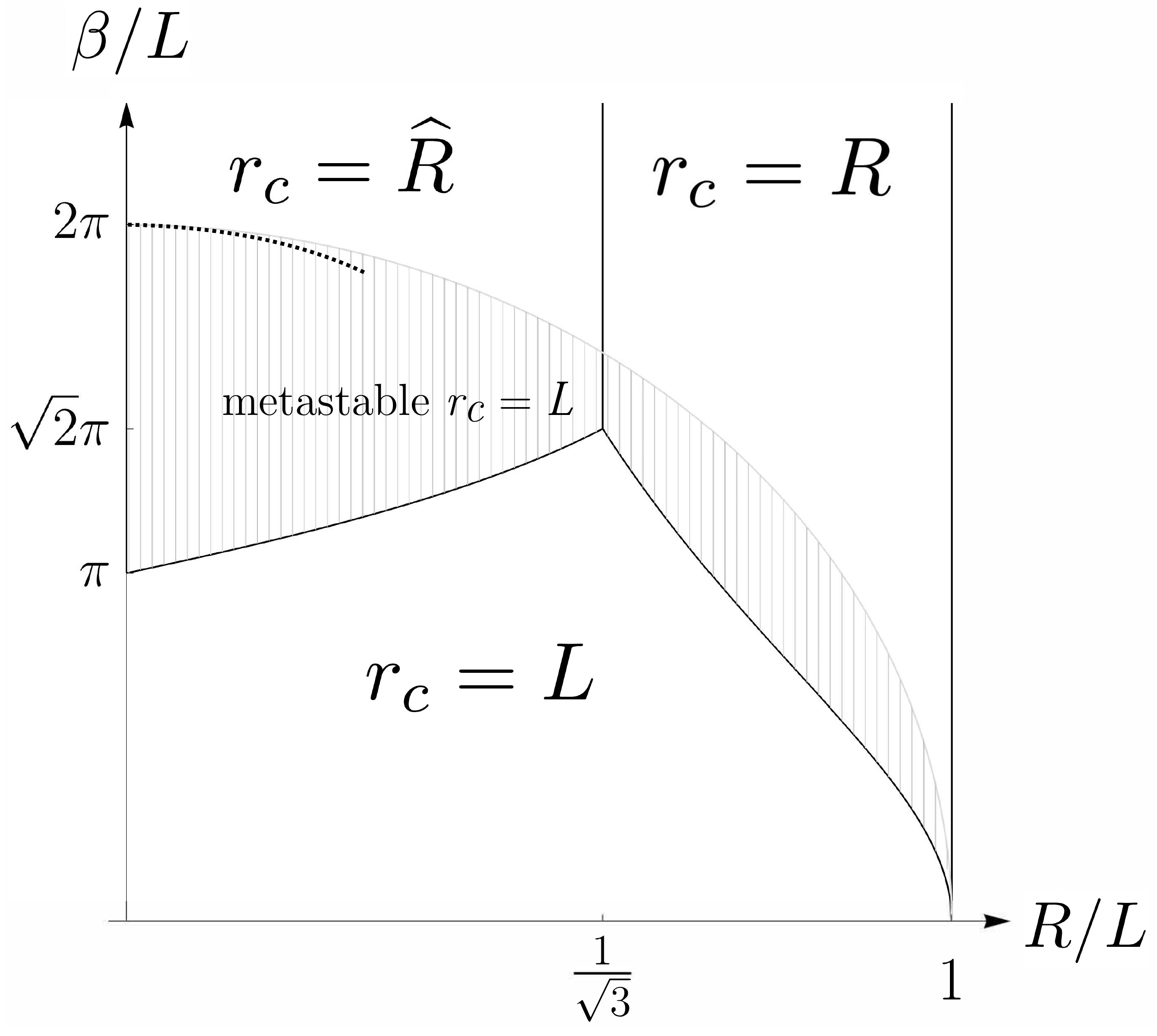}
    \caption{\small  The phase diagram for cosmological horizon patches in $d=4$ spacetime dimensions, with boundary size $R$ and inverse temperature $\beta$ (figure adapted from \cite{Banihashemi:2022jys}). The temperature on the  upper (semicircular) boundary of the metastable $r_c=L$ region  is  the GH Tolman temperature at $R$. 
    In the metastable region the 
    configuration with $r_c=L$ is a local (endpoint) minimum of the action, 
    while the absolute (endpoint) minimum has either
    $r_c = R$ or $r_c= \widehat R$, where $\widehat{R}$
    is the radius of the other horizon when one horizon
     has radius $R$.
    The dotted path describes the $R\to 0$ sequence of metastable de Sitter patches considered in the main body of this paper. Not shown are the black hole and thermal de Sitter phases, which have higher action along the dashed path~\cite{Banihashemi:2022jys}. }
    \label{fig:phasediag}
\end{figure}

In \cite{Banihashemi:2022jys} the thermal instability of the 
cosmological system was inferred from the fact that  the cosmological horizon solution is a local maximum (instead of a minimum) of the action as a function of the mass parameter in the spatial SdS line element.  If negative Schwarzschild-de Sitter mass parameters are allowed in the solutions to the constraints, the action is unbounded below, corresponding to arbitrarily large horizon size. However, it was argued in \cite{Banihashemi:2022jys} that the York boundary conditions defining the ensemble should represent the interface with a physically sensible reservoir, and that negative mass parameter would correspond to a reservoir whose energy is unbounded below, hence that configurations with negative mass parameters should be excluded. If this is done, the minimum of the action for configurations with a cosmological horizon lies at one of the two endpoints of the allowed configuration space. For a given boundary size, at sufficiently high temperatures, the endpoint with minimum action has a horizon radius equal to the de Sitter radius (corresponding to zero mass parameter), whereas at lower temperatures the minimum lies at the other endpoint which is a configuration in which the boundary coincides with 
  a black hole horizon for $R<L/\sqrt{3}$ and a cosmological horizon for $R>L/\sqrt{3}$.
One can choose a path through the $(R,T)$ parameter space, 
with $T$ greater than the GH Tolman temperature at $R$ and
ending at $(R=0, T=\hbar/2\pi L)$, 
along which the configuration with de Sitter radius horizon is metastable since it lies at a local but not global minimum of the action. 
 Figure \ref{fig:phasediag}, taken from \cite{Banihashemi:2022jys}, shows the 
phase diagram of the system, on which we have superimposed a path of the
sort just described. 
In the main text we consider such a path in the limit that
$T$ approaches the GH Tolman temperature at $R$, for which the 
configurations are marginally stable, so that the first law for variations
away from an on-shell configuration may be applied.

 \bibliographystyle{JHEP}
 \bibliography{dSrefs}

\end{document}